\documentclass[aps, prd, amsmath, floats, floatfix, twocolumn, nofootinbib, superscriptaddress]{revtex4-1}
\usepackage{graphicx}
\usepackage{color}
\usepackage{latexsym}

\newcommand{\be}{\begin{eqnarray}}
\newcommand{\ee}{\end{eqnarray}}
\newcommand{\rar}{\rightarrow}

\begin{document}

\title{A parametrization to test black hole candidates with the spectrum of thin disks}

\author{Nan Lin}
\affiliation{Center for Field Theory and Particle Physics and Department of Physics, Fudan University, 200433 Shanghai, China}

\author{Naoki Tsukamoto}
\affiliation{Center for Field Theory and Particle Physics and Department of Physics, Fudan University, 200433 Shanghai, China}

\author{M. Ghasemi-Nodehi}
\affiliation{Center for Field Theory and Particle Physics and Department of Physics, Fudan University, 200433 Shanghai, China}

\author{Cosimo Bambi}
\email[Corresponding author: ]{bambi@fudan.edu.cn}
\affiliation{Center for Field Theory and Particle Physics and Department of Physics, Fudan University, 200433 Shanghai, China}
\affiliation{Theoretical Astrophysics, Eberhard-Karls Universit\"at T\"ubingen, 72076 T\"ubingen, Germany}

\date{\today}

\begin{abstract}
We discuss a parametrization to describe possible deviations from the Kerr metric and test astrophysical black hole candidates with electromagnetic radiation. Our metric is a very simple generalization of the Kerr solution with two main properties: $i)$ the phenomenology is quite rich and, for example, it can describe black holes with high Novikov-Thorne radiative efficiency or black holes of very small size; $ii)$ it is suitable for the numerical calculations required to study the spectrum of thin disks. The latter point is our principal motivation to study such a kind of parametrization, because in the analysis of real data there are usually several parameters to fit and the problem with current non-Kerr metrics is that the calculation times are too long.
\end{abstract}

\maketitle


\section{Introduction}

Astrophysical black hole candidates are dark and compact objects that can be naturally interpreted as black holes and they may be something else only in the presence of new physics. Stellar-mass black hole candidates are compact objects in X-ray binaries with a mass exceeding the maximum mass for a neutron star~\cite{rr}. Supermassive black hole candidates are the huge compact bodies at the center of every normal galaxy and they turn out to be too massive, compact, and too old to be a cluster of neutron stars~\cite{maoz}. The non-detection of thermal radiation from the surface of these objects is also consistent with the idea that they do not have a surface but an event horizon~\cite{h1,h2}.

According to general relativity, the spacetime metric around black hole candidates should be well described by the Kerr solution. Initial deviations from the Kerr metric are quickly radiated away through the emission of gravitational waves~\cite{k1}. The equilibrium electric charge is completely negligible for macroscopic objects~\cite{k2}. The accretion disk is usually many orders of magnitude smaller than the central black hole candidate and it cannot appreciably change the geometry of the spacetime~\cite{k3}.

The Kerr black hole hypothesis entirely relies on the validity of general relativity and there is no clear observational confirmation that the spacetime geometry around black hole candidates is described by the Kerr solution. Moreover, general relativity has been tested only for weak gravitational fields and it is not guaranteed that its predictions still hold in the strong gravity regime.

In the past few years, there have been a significant work to study how present and future observational facilities could test black hole candidates, see e.g. Refs.~\cite{rev1,rev2}. The most common approach is to use a method similar to the Parametrized Post-Newtonian (PPN) formalism~\cite{ppn68}, in which one wants to test the Schwarzschild solution in the weak field limit. In the case of black hole candidates, one employes a metric that is parametrized by a number of deformation parameters capable of describing possible deviations from the Kerr background. The deformation parameters are free quantities to be determined by observations, and {\it a posteriori} one can check whether astronomical data demand vanishing deformation parameters, as it is required by the Kerr solution.

In the literature there is already a number of parametrizations suitable to test black hole candidates~\cite{p1,p2,p3,p4,p5,p6}. Each proposal has its own advantages and disadvantages. However, in the analysis of real data it is necessary to calculate a large number of spectra for different values of the model parameters in order to find the best fit and measure the model parameters. Typical calculations require the determination of the point of the emission on the accretion disk and of its redshift factor~\cite{tt1,tt2,tt3,tt4,tt5,tt6,tt7}. The Kerr metric has some nice properties, and eventually these calculations can be done in a reasonable time with current computational facilities. The non-Kerr metrics used to test black hole candidates do not have such nice properties and this becomes an issue when we want to measure the deformation parameters with real data.

In the present paper, we discuss a parametrization suitable for numerical calculations involving the electromagnetic spectrum of thin disks. This is our main motivation, and in particular we have in mind the continuum-fitting and the iron line methods~\cite{tt1,tt2,tt3,tt4,tt5,tt6,tt7,bjs15}. We are thus interested in a metric with properties similar to the Kerr solution, and we do not look for a very general black hole metric. Our metric has the Carter constant and therefore the equations of motion are separable and of first order. More importantly, the motion along the $\theta$-direction is like in Kerr metric and it can be reduced to an elliptic integral, while the motion along the $r$-direction can be reduced to a hyper-elliptic integral. In general, this is probably the best we can have from the point of view of the accuracy and the speed of the calculations. We also note that our metric has a quite rich phenomenology. For instance, it can describe black holes with a very high Novikov-Thorne radiative efficiency, which is not the case for most (if not all) parametrizations already proposed in the literature. Such a property is quite useful when we have fast-rotating objects, which are the best candidates to test the Kerr paradigm. Another feature, which is usually absent in the other parametrization, is that our black holes can be very small. It is also worth noting that our metric has no curvature singularities outside of the event horizon.

\section{Metric}

As in the other parametrizations discussed in the literature~\cite{p1,p2,p3,p4,p5,p6}, even our choice is necessarily {\it ad hoc} and it can only be motivated by our requirements, which are determined by the specific use we have in mind. The metric must clearly includes the Kerr solution as a special case. We want that there is the Carter constant, so that it is not necessary to solve the geodesic equations but the equations of motion are separable and of first order. To do this, we write the Kerr metric in Boyer-Lindquist coordinates and we promote the constant $M$ to some functions $m_i (r)$ that depend on the radial coordinate only. The line element reads
\be\label{eq-m}
ds^2 &=& - \left(1 - \frac{2 m_1 r}{\Sigma}\right) dt^2
- \frac{4 a m_1 r \sin^2\theta}{\Sigma} dtd\phi \nonumber\\
&& + \frac{\Sigma}{\Delta_2} dr^2 + \Sigma d\theta^2 \nonumber\\
&& + \left(r^2 + a^2 + \frac{2 a^2 m_1 r \sin^2\theta}{\Sigma}\right) \sin^2\theta d\phi^2 \, ,
\ee
where  $\Sigma = r^2 + a^2 \cos^2\theta$, $\Delta_2 = r^2 - 2 m_2 r + a^2$, $m_1 = m_1(r)$, and $m_2 = m_2(r)$. In $g_{tt}$, $g_{t\phi}$, and $g_{\phi\phi}$, $M$ has been replaced by the same function $m_1$, because otherwise we lose the separability of the equations of motion (see the next section). The mass $M$ in $g_{rr}$ has been replaced by the function $m_2$, which (in general) may be different by $m_1$ without affecting our requirement.

In the metric in Eq.~(\ref{eq-m}), the radius of the event horizon, $R_{\rm H}$, is given by the largest root in $g^{rr} = 0$, namely $\Delta_2 = 0$, where only $m_2$ is involved, not $m_1$. On the other hand, the Killing horizon is given by the largest root in
\be
g_{tt} g_{\phi\phi} - g_{t\phi}^2 = 0 \, .
\ee
If $m_1 = m_2$, an extension of the rigidity theorem holds and event horizon and Killing horizon coincide. In the general case, with $m_1 \neq m_2$, this may not be true.

We note that the metric in Eq.~(\ref{eq-m}) reduces to the Kerr-Newman solution when
\be
m_1 = m_2 = M - \frac{Q^2}{2 r} \, ,
\ee
where $Q$ is the electric charge of the black hole. Our metric can also describe a large class of quantum gravity inspired black hole solutions~\cite{nc1,nc2,nc3,nc4,wngt} and some regular black holes~\cite{bardeen,regular}. For instance, in the non-commutative inspired black holes of Refs.~\cite{nc1,nc2,nc3,nc4}, one has
\be
m_1(r) = m_2(r) = \frac{\gamma (3/2;r^2/4l_0^2)}{\Gamma(3/2)} \, M \, ,
\ee
where $\gamma (3/2;r^2/4l_0^2)$ is the lower incomplete Gamma function, $\Gamma(3/2) = \sqrt{\pi}/2$ is the Gamma function at $3/2$, and $l_0$ is the non-commutativity length scale of the theory. In the weakly non-local theories of gravity of Ref.~\cite{wngt}, the black hole solutions have 
\be
&& m_1(r) = m_2(r) = \nonumber\\
&& \;\;\; = \frac{2M}{\pi} \int_0^r dx \, x^2
\int^\infty_0 dk \, k^2 \frac{\sin kr}{kr} 
V\left(-\frac{k^2}{\Lambda^2}\right) \, ,
\ee
where $V$ is the model-dependent form factor and $\Lambda$ is the scale of the theory. The Bardeen metric~\cite{bardeen,regular} has
\be
m_1(r) = m_2(r) = \frac{r^3}{\left(r^2 + g^2\right)^{3/2}} \, M \, .
\ee
If the Bardeen solution is derived from Einstein gravity coupled to a non-linear electrodynamics field~\cite{bardeen2}, $g$ is the magnetic charge of the black hole. In all these examples, we always have $m_1 = m_2$, but in what follows we will consider the general case without this condition when it is not required the exact expression of $m_1$ and $m_2$.

Let us write $m_1$ and $m_2$ in the following form
\be\label{eq-m1m2}
m_i = M \sum_{k=0}^{\infty} a_{ik} \left(\frac{M}{r}\right)^k \, .
\ee
In the weak field regime, $M/r \ll 1$ and can be used as an expansion parameter. The metric coefficient $g_{tt}$ and $g_{rr}$ become
\be
g_{tt} &=& - \left[ 1 - a_{10} \frac{2 M}{r} - a_{11} \frac{2 M^2}{r^2} + . . . \right] \, , \\
g_{rr} &=& 1 + a_{20} \frac{2 M}{r} + . . . \, .
\ee
When cast in Schwarzschild coordinates, the PPN metric reduces to
\be
g_{tt} &=& - \left[ 1 - \frac{2 M}{r} + \left( \beta_{\rm PPN} - \gamma_{\rm PPN} \right) \frac{2 M^2}{r^2} + . . . \right] \, , \\
g_{rr} &=& 1 + \gamma \frac{2 M}{r} + . . . \, ,
\ee
and Solar System experiments require that $\beta_{\rm PPN}$ and $\gamma_{\rm PPN}$ are 1 with an accuracy at the level of $10^{-5} - 10^{-4}$~\cite{will}. Within our parametrization and $m_1$ and $m_2$ given by Eq.~(\ref{eq-m1m2}), we can always choose $a_{10} = 1$ (if $a_{10} \neq 1$, we just redefine $M$). $a_{11} \approx 0$ and $a_{20} \approx 1$ are constrained by Solar System experiments. The first unconstrained coefficients are thus $a_{12}$ and $a_{21}$.

In the next sections, we will focus the attention on the following choice of $m_1$ and $m_2$:
\be\label{eq-Mab}
m_1 = m_2 = M \left( 1 + \alpha \frac{M^2}{r^2} + \beta \frac{M^3}{r^3} \right) \, ,
\ee
where $\alpha$ and $\beta$ are the two deformation parameters of our metric. This choice is obtained from the truncation of the general expression in Eq.~(\ref{eq-m1m2}); that is, we consider the two leading order terms without Solar System constraints and we neglect higher order corrections. The Kerr solution is recovered when $\alpha = \beta = 0$. With the choice in Eq.~(\ref{eq-Mab}), there are no naked singularities in the region outside of the black hole. In the Appendix, we report the expressions of the invariants $R$, $R^{\mu\nu} R_{\mu\nu}$, and $R^{\mu\nu\rho\sigma} R_{\mu\nu\rho\sigma}$. These invariants are everywhere regular except at $r=0$.

Our final goal is to have a parametrization suitable for the numerical calculations necessary in the study of the electromagnetic spectrum of thin disks. The background metric enters the following calculations:
\begin{enumerate}
\item The motion of the particle in the disk. In the Novikov-Thorne model~\cite{nt-model}, the disk is on the equatorial plane orthogonal to the black hole spin, and the particles of the accretion disk follow nearly geodesic circular orbits on the equatorial plane.
\item The photon trajectories from the emission point in the disk to the detection point at infinity. Actually, it is not strictly necessary to compute the exact photon trajectories, but it is necessary to connect the emission point on the disk to that on the image plane of the distant observer~\cite{c75,srr}.
\end{enumerate}

\section{Motion of massive particles in the equatorial plane}

In the calculations of the motion of the particles in the disk, only the metric coefficient $g_{tt}$, $g_{t\phi}$, and $g_{\phi\phi}$ are involved. So we need to specify $m_1$, which will be assumed to have the form in Eq.~(\ref{eq-Mab}), while $m_2$ is actually irrelevant in this part.

The angular velocity of equatorial circular orbits is~\cite{teu72,tt1}
\be
\Omega_\pm &=& \frac{- \left(\partial_r g_{t\phi}\right) \pm \sqrt{\left(\partial_r g_{t\phi}\right)^2 - \left(\partial_r g_{tt}\right) \left(\partial_r g_{\phi\phi}\right)}}{\partial_r g_{\phi\phi}} \nonumber\\
&=& \frac{\tilde{m}_1^{1/2}}{r^{3/2} \pm a \tilde{m}_1^{1/2}} \, ,
\ee
where here and in the next formulas the upper sign refer to corotating orbits, while the lower sign to counterrotating orbits. Moreover, we have introduced $\tilde{m}_1$, which is defined by
\be
\tilde{m}_1 = M \left( 1 + 3 \alpha \frac{M^2}{r^2} + 4 \beta \frac{M^3}{r^3} \right)
\ee
When $\alpha = \beta = 0$, we clearly recover the well-known Kerr result~\cite{teu72}
\be
\Omega_\pm^{\rm Kerr} = 
\frac{M^{1/2}}{r^{3/2} \pm a M^{1/2}} \, .
\ee

Following the standard prescription to find the equatorial circular orbits~\cite{teu72,tt1}, the specific energy and specific angular momentum of the particles of the gas are
\be\label{eq-e}
E &=& \frac{r^{3/2} - 2 m_1 r^{1/2} \pm a \tilde{m}_1^{1/2}}{r^{3/4} 
\sqrt{r^{3/2} - 2 m_1 r^{1/2} - \tilde{m}_1 r^{1/2} \pm 2 a \tilde{m}_1^{1/2}}} \, , \\
\label{eq-l}
L_z &=& \pm \frac{\tilde{m}_1^{1/2} 
\left(r^2 \mp 2 a m_1^{1/2} \tilde{m}_1^{-1/2} r^{1/2} + a^2\right)}{r^{3/4} 
\sqrt{r^{3/2} - 2 m_1 r^{1/2} - \tilde{m}_1 r^{1/2} \pm 2 a \tilde{m}_1^{1/2}}} \, .
\nonumber\\
\ee
We recover the correct Kerr limit for $\alpha = \beta = 0$
\be
E &=& \frac{r^{3/2} - 2 M r^{1/2} \pm a M^{1/2}}{r^{3/4} 
\sqrt{r^{3/2} - 3 M r^{1/2} \pm 2 a M^{1/2}}} \, , \\
L_z &=& \pm \frac{M^{1/2} \left(r^2 \mp 2 a M^{1/2} r^{1/2} + a^2\right)}{r^{3/4} 
\sqrt{r^{3/2} - 3 M r^{1/2} \pm 2 a M^{1/2}}} \, .
\ee

When
\be\label{eq-po}
r^{3/2} - 2 m_1 r^{1/2} - \tilde{m}_1 r^{1/2} \pm 2 a \tilde{m}_1^{1/2} = 0 \, ,
\ee
the denominator in Eqs.~(\ref{eq-e}) and (\ref{eq-l}) vanishes and the particle has infinite energy. Eq.~(\ref{eq-po}) defines the radius of the photon orbit, $R_{\rm photon}$, which is the minimum radius for circular orbits (at smaller radii, there are no circular orbits).

In the spectrum of thin disks, the inner edge of the disk plays a crucial role and it is eventually the true key-ingredient in both the continuum-fitting and the iron line methods. In the Novikov-Thorne model, the inner edge of the disk is at the innermost stable circular orbit (ISCO). In general, one has to check the orbital stability along both the radial and the vertical directions~\cite{tt1,bb2}. In our spacetime we have checked that the ISCO radius is only determined by the orbital stability along the radial direction (like in the Kerr metric), at least for not too large values of the deformation parameters. In this case, the ISCO radius corresponds to the minimum of the specific energy 
\be
\frac{dE}{dr} = 0 \Rightarrow r = R_{\rm ISCO} \, .
\ee
Unfortunately, it seems there is not a compact analytic expression for $R_{\rm ISCO}$ as in the Kerr metric.

The so-called Novikov-Thorne radiative efficiency, which is the actual quantities measured by the continuum-fitting method~\cite{tt4}, is
\be
\eta_{\rm NT} = 1 - E_{\rm ISCO} \, ,
\ee
where $E_{\rm ISCO}$ is the specific energy of a test particle at the ISCO radius, namely Eq.~(\ref{eq-e}) evaluated at $r = R_{\rm ISCO}$

The 4-velocity of a particle in the disk is given by $u^\mu_{\rm e} = (u^t_{\rm e}, 0, 0, u^\phi_{\rm e})$, where $u^\phi_{\rm e} = \Omega u^t_{\rm e}$ by definition of $\Omega$. From the normalization condition $g_{\mu\nu} u^\mu_{\rm e} u^\nu_{\rm e} = -1$, we get the expression for $u^t_{\rm e}$
\be
u^t_{\rm e} &=& \frac{1}{\sqrt{-g_{tt} - 2g_{t\phi}\Omega - g_{\phi\phi}\Omega^2}}
\nonumber\\
&=&
\frac{r^{3/2}_{\rm e} + a \tilde{m}_1^{1/2}}{r^{1/2}_{\rm e} 
\sqrt{r^2_{\rm e} - 2 m_1 r_{\rm e} - \tilde{m}_1 
r_{\rm e} + 2 a \tilde{m}_1^{1/2} r^{1/2}_{\rm e}}} \, ,
\ee
which reduces to the correct Kerr case for $\alpha = \beta = 0$
\be
u^t_{\rm e} = \frac{r^{3/2}_{\rm e} + a M^{1/2}}{r^{1/2}_{\rm e} \sqrt{r^2_{\rm e} - 3 M r_{\rm e} + 2 a M^{1/2} r^{1/2}_{\rm e}}} \, .
\ee

The 4-velocity of the particles in the disk is necessary to calculate another important quantity in the calculation of the spectrum of thin disks, namely the redshift factor $g$
\be
g &=& \frac{\nu_{\rm o}}{\nu_{\rm e}}
= \frac{u^\mu_{\rm o} k_\mu}{u^\nu_{\rm e} k_\nu} 
= \frac{\sqrt{-g_{tt} - 2g_{t\phi}\Omega - g_{\phi\phi}\Omega^2}}{1 - \xi \Omega} \, ,
\ee
where $\nu_{\rm o}$ and $\nu_{\rm e}$ are, respectively, the photon frequency as measured by the distant observer and the emitter, $u^\mu_{\rm o} = (1,0,0,0)$ is the 4-velocity of the observer, $k^\mu$ is the 4-momentum of the photon, and $\xi = - k_\phi/k_t$ is a constant of motion along the photon trajectory (as a consequence of the fact the spacetime is stationary and axisymmetric). Within our parametrization, we find (for $a \ge 0$)
\be
g &=& \frac{r^{1/2}_{\rm e} \sqrt{r^2_{\rm e} - 2 m_1 r_{\rm e} - \tilde{m}_1 r_{\rm e} 
+ 2 a \tilde{m}_1^{1/2} r^{1/2}_{\rm e}}}{r^{3/2}_{\rm e} 
+ a \tilde{m}_1^{1/2} - \tilde{m}_1^{1/2} \xi} \, .
\ee
The correct Kerr limit is again recovered for $\alpha = \beta = 0$
\be
g = \frac{r_{\rm e}^{1/2} \sqrt{r_{\rm e}^2 - 3 M r_{\rm e} 
+ 2 a M^{1/2} r^{1/2}_{\rm e}}}{r^{3/2}_{\rm e} + a M^{1/2} - M^{1/2} \xi} \, .
\ee
In the end, we have $g = g(r_{\rm e},\xi)$. Since $r_{\rm e} = r_{\rm e}(\xi,q)$, where $q^2={\mathcal Q}/E^2$ and ${\mathcal Q}$ is the Carter constant (see Section~\ref{s-sss}), eventually we have~\cite{c75}
\be\label{eq-g-g}
g = g (\xi, q) \, .
\ee

\begin{figure*}
\begin{center}
\includegraphics[type=pdf,ext=.pdf,read=.pdf,width=8.5cm]{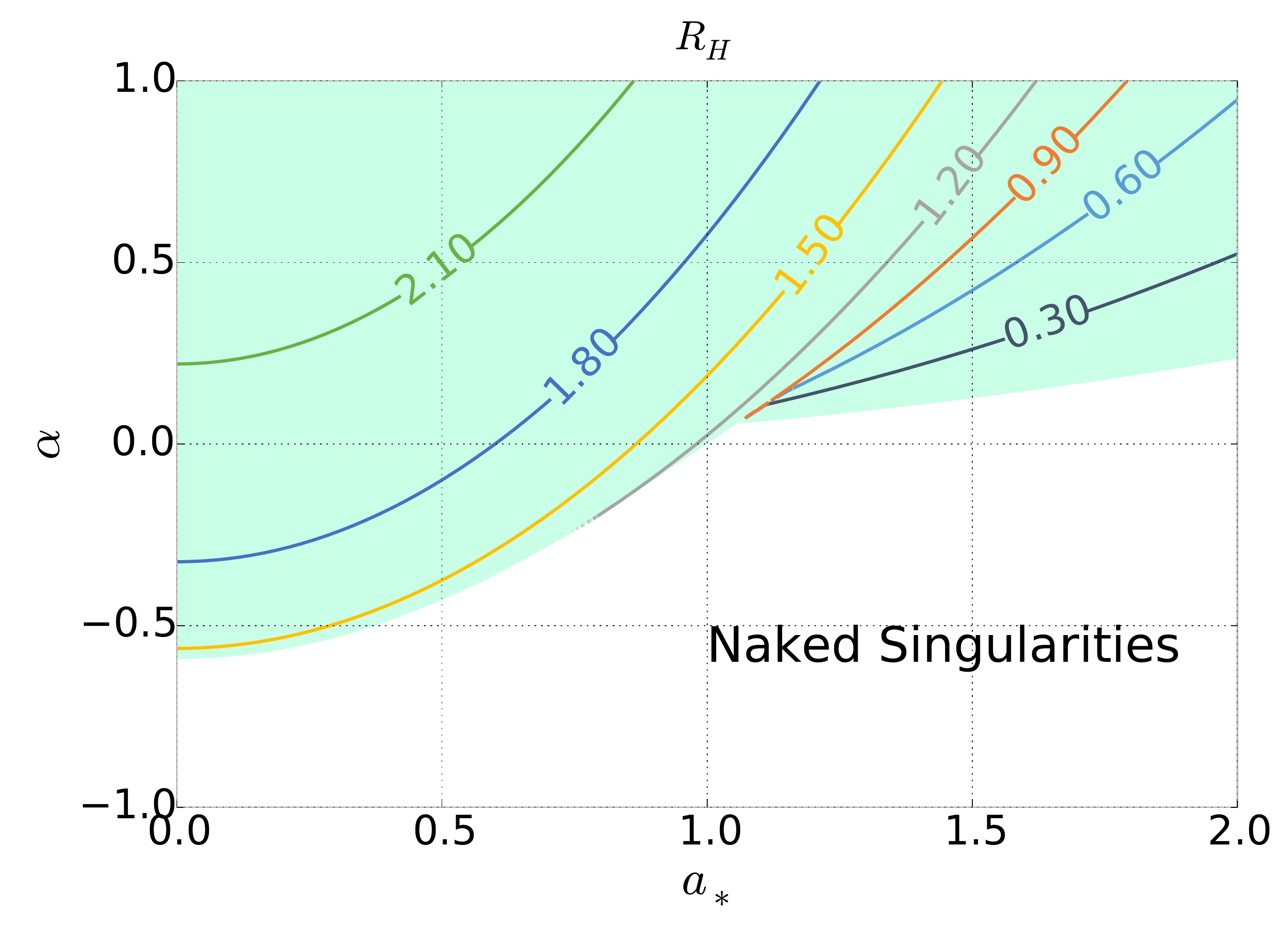}
\hspace{0.3cm}
\includegraphics[type=pdf,ext=.pdf,read=.pdf,width=8.5cm]{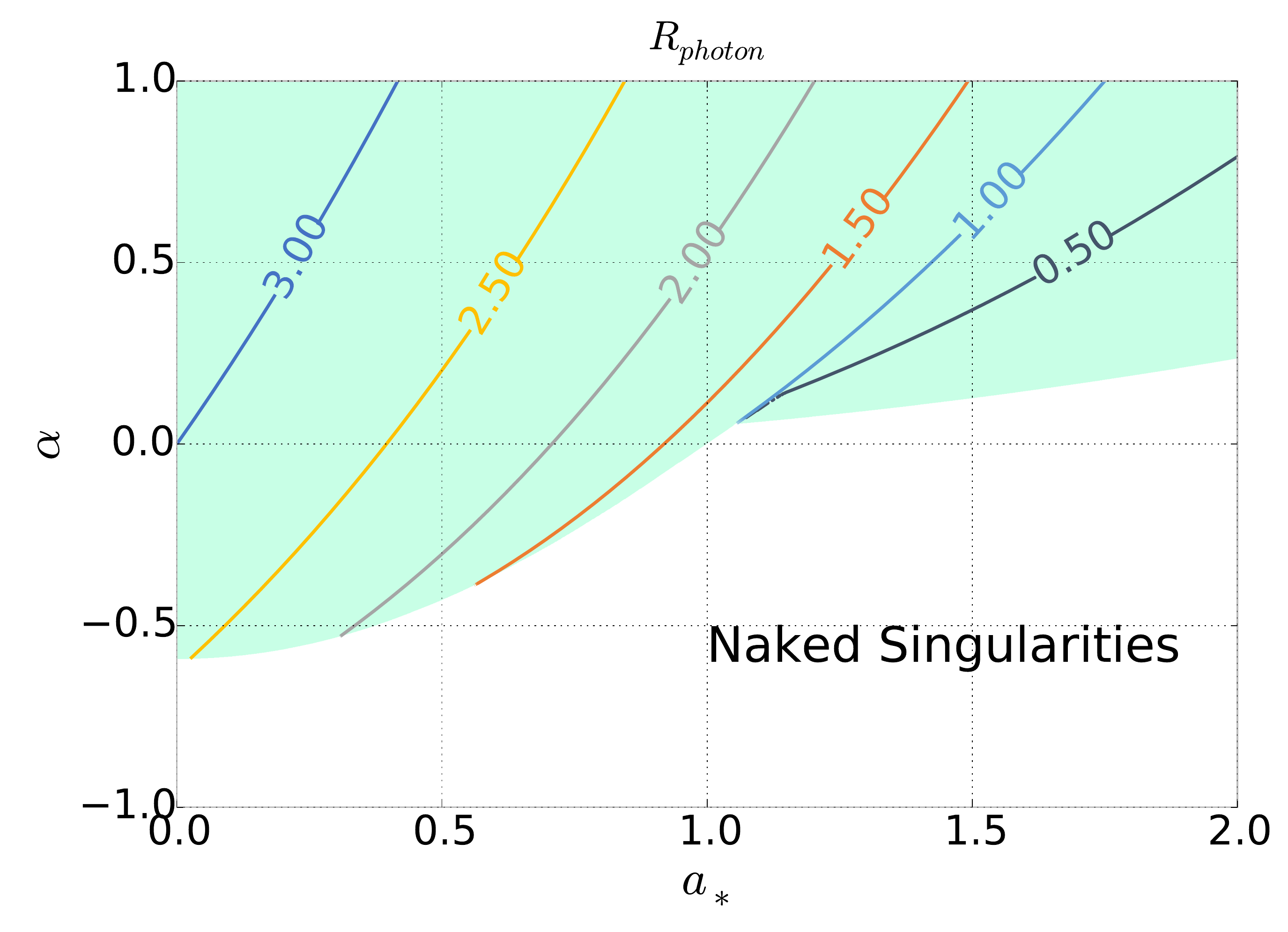} \\ 
\vspace{0.5cm}
\includegraphics[type=pdf,ext=.pdf,read=.pdf,width=8.5cm]{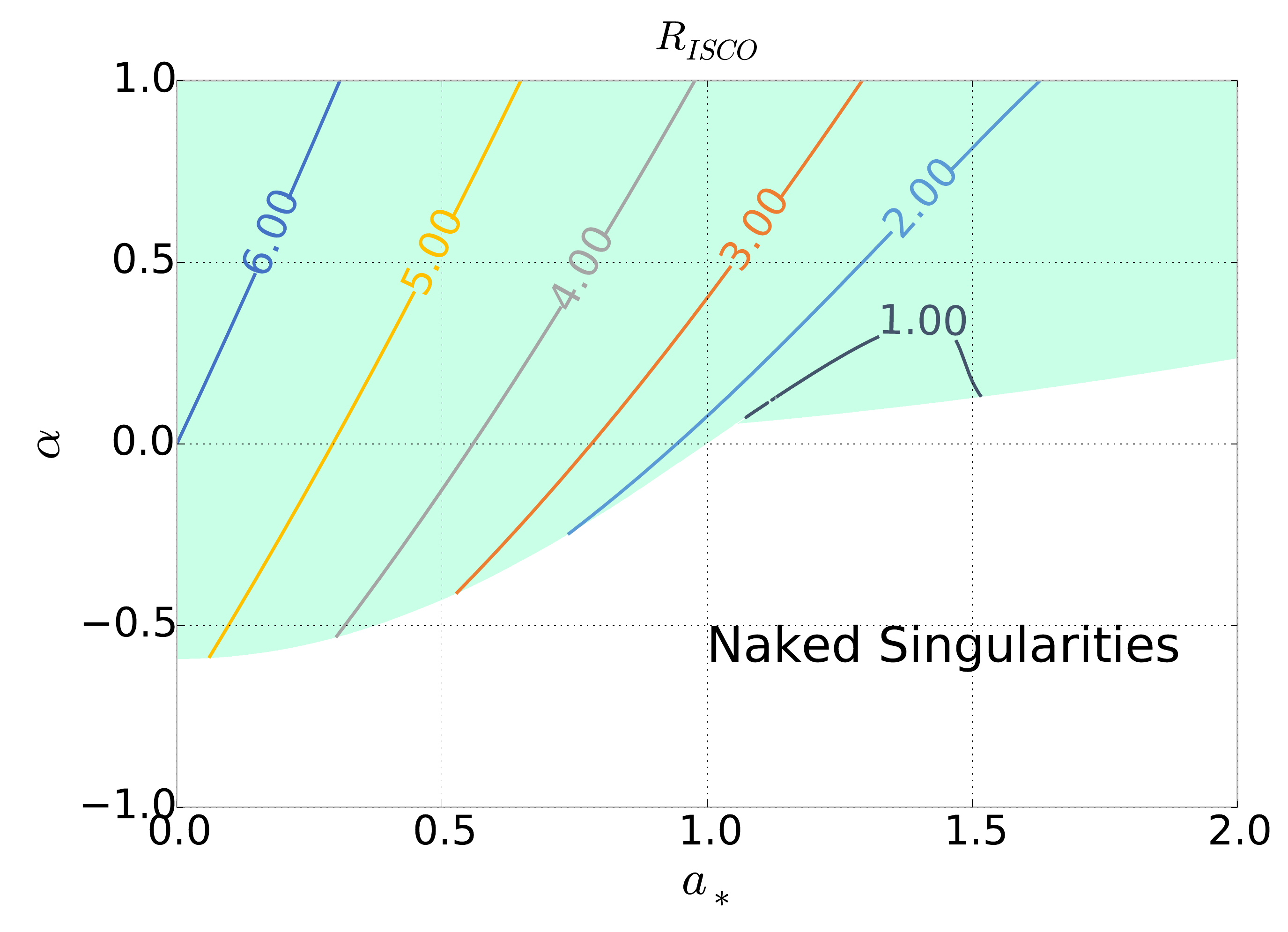}
\hspace{0.3cm}
\includegraphics[type=pdf,ext=.pdf,read=.pdf,width=8.5cm]{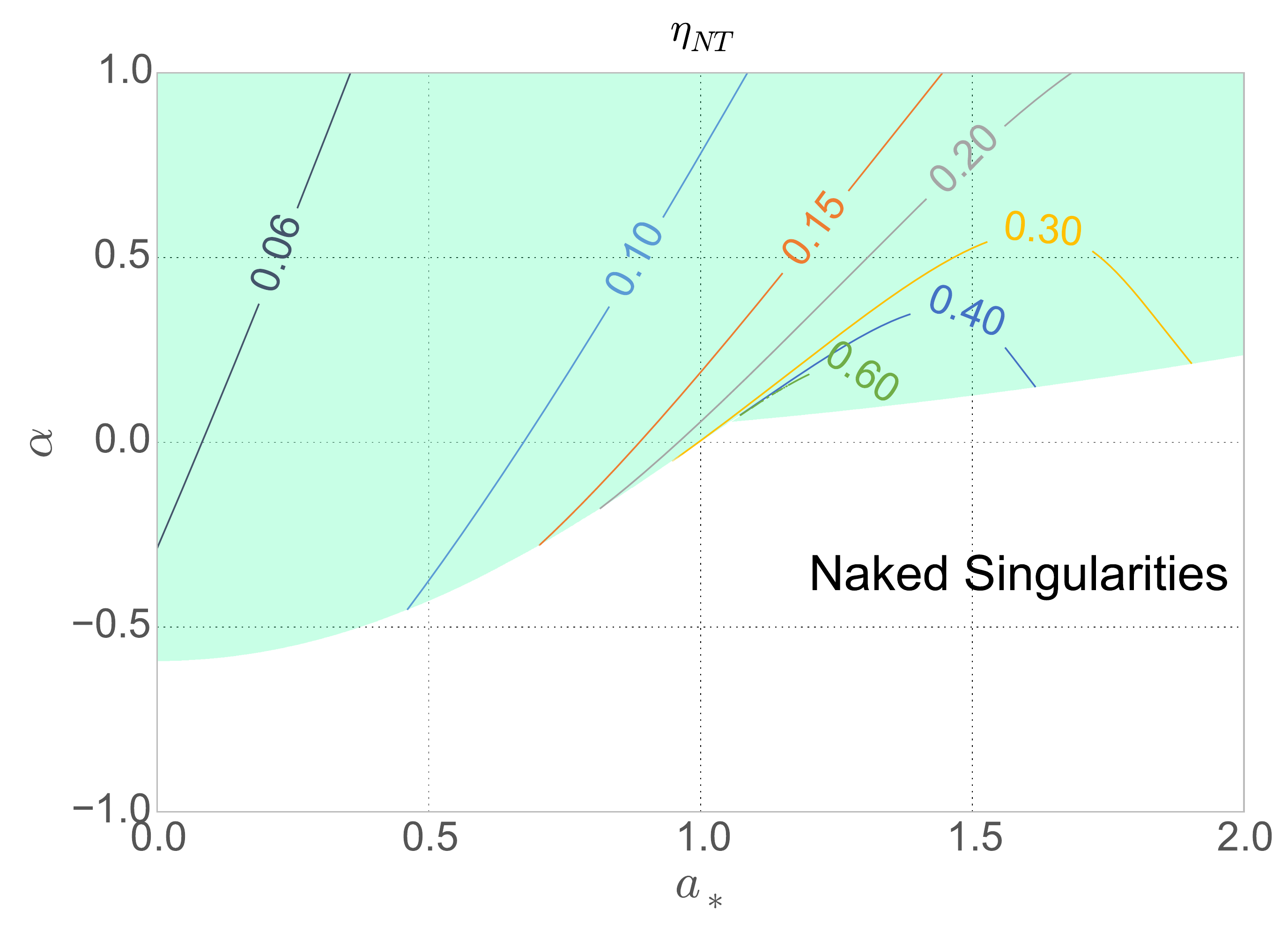}
\end{center}
\caption{Contour levels of the radius of the event horizon $R_{\rm H}$ (top left panel), of the photon radius $R_{\rm photon}$ (top right panel), of the ISCO radius $R_{\rm ISCO}$ (bottom left panel), and of the Novikov-Thorne radiative efficiency $\eta_{\rm NT}$ (bottom right panel) for $\beta = 0$. $m_1$ and $m_2$ are given by Eq.~(\ref{eq-Mab}). The spacetimes in the white region have no black hole but a naked singularity and they have not been studied. $a_* = a/M$ is the dimensionless spin parameter. $R_{\rm H}$, $R_{\rm photon}$, and $R_{\rm ISCO}$ in units with $M=1$. See the text for more details.}
\label{fig-b0.0}
\end{figure*}

\begin{figure*}
\begin{center}
\includegraphics[type=pdf,ext=.pdf,read=.pdf,width=8.5cm]{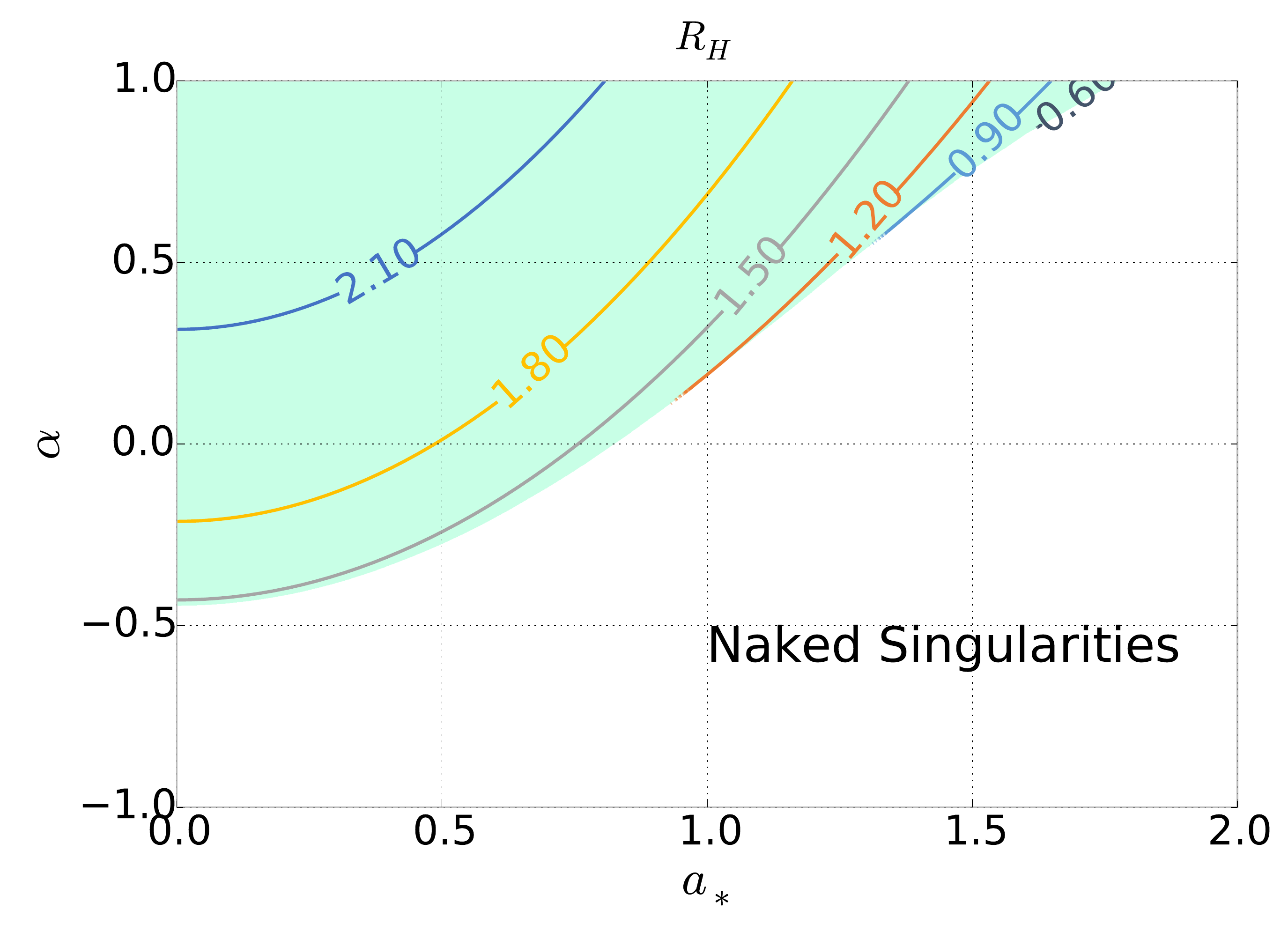}
\hspace{0.3cm}
\includegraphics[type=pdf,ext=.pdf,read=.pdf,width=8.5cm]{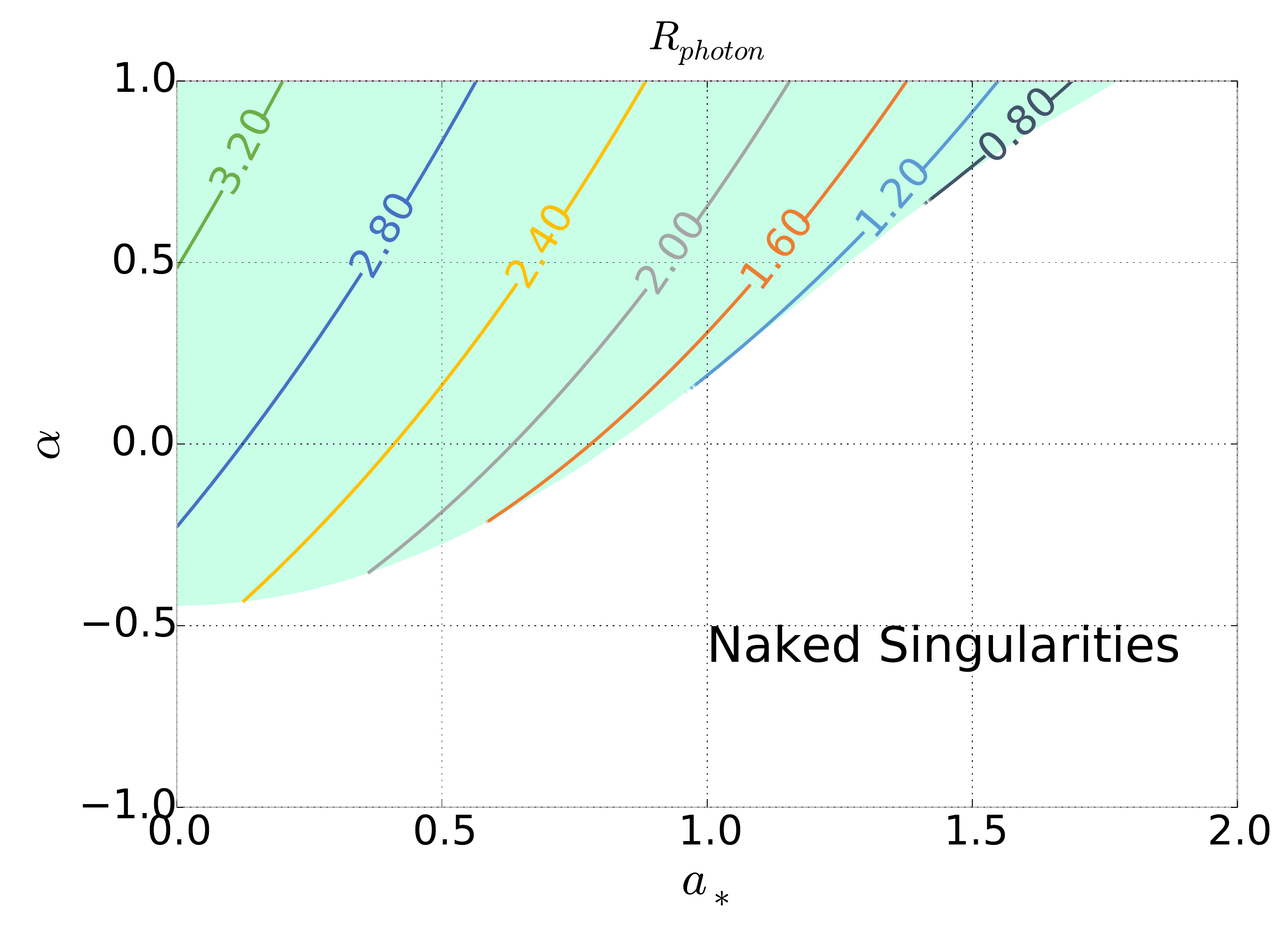} \\ 
\vspace{0.5cm}
\includegraphics[type=pdf,ext=.pdf,read=.pdf,width=8.5cm]{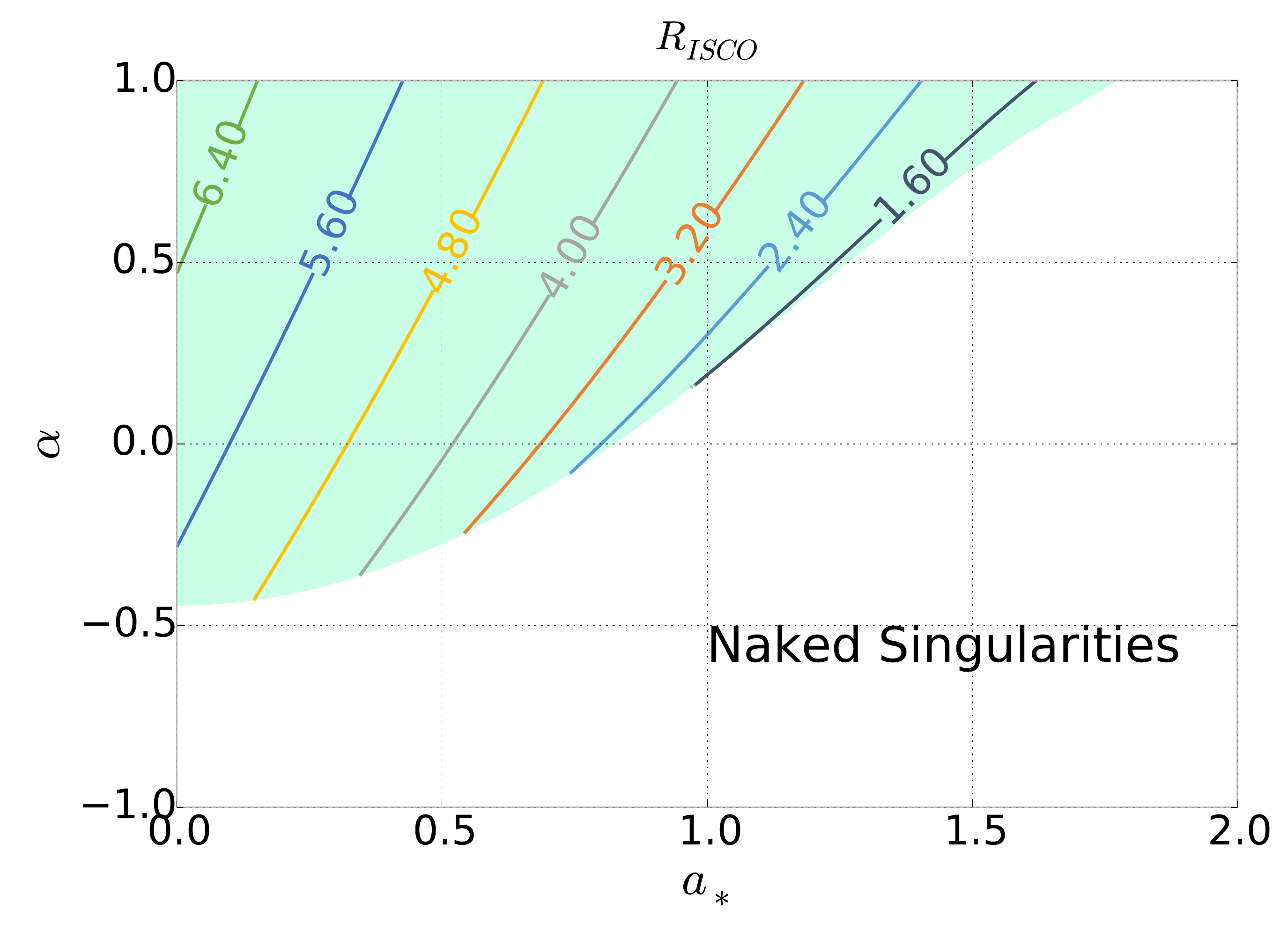}
\hspace{0.3cm}
\includegraphics[type=pdf,ext=.pdf,read=.pdf,width=8.5cm]{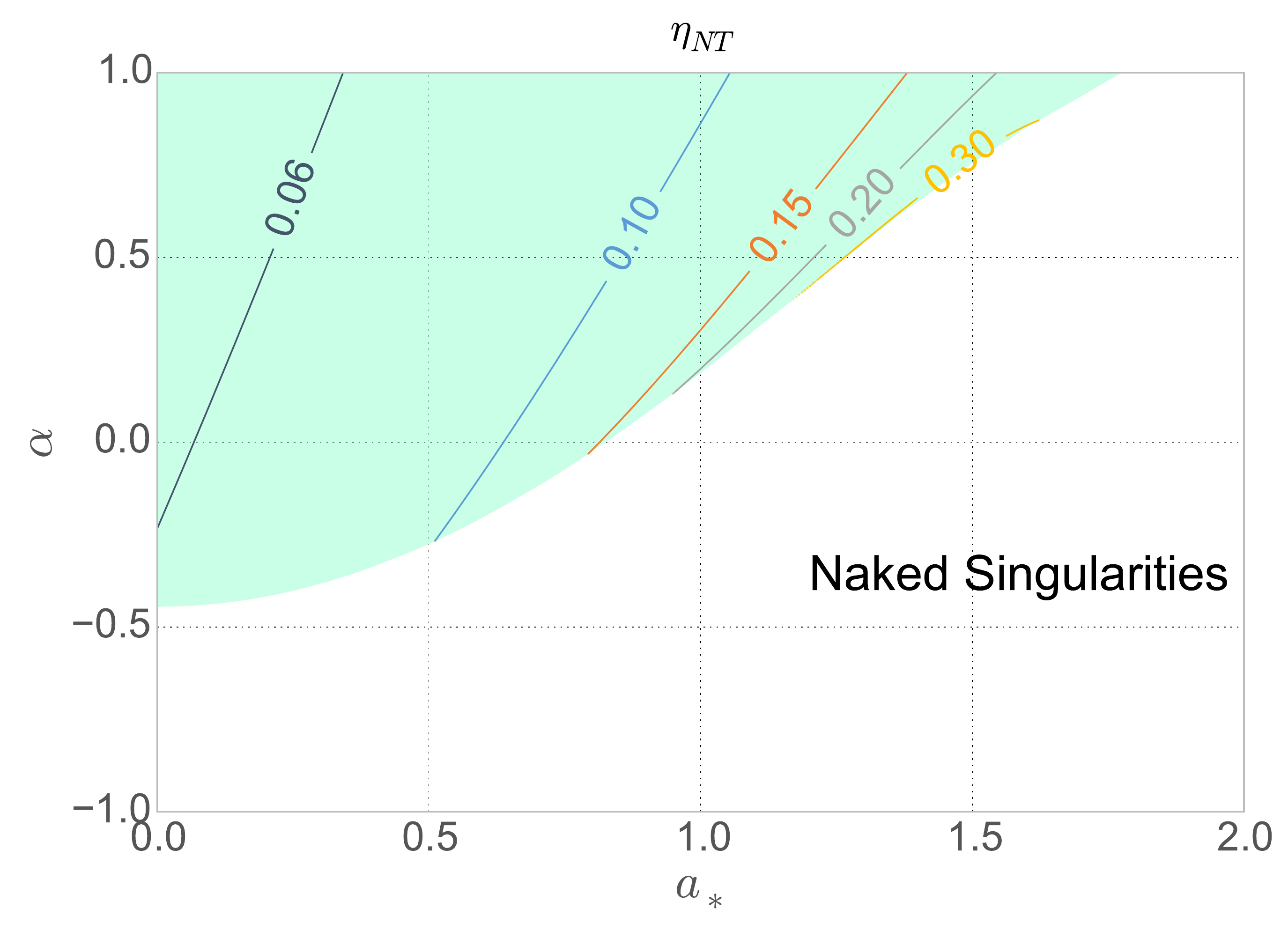}
\end{center}
\caption{As in Fig.~\ref{fig-b0.0} for $\beta = -0.2$.}
\label{fig-b-0.2}
\end{figure*}

\begin{figure*}
\begin{center}
\includegraphics[type=pdf,ext=.pdf,read=.pdf,width=8.5cm]{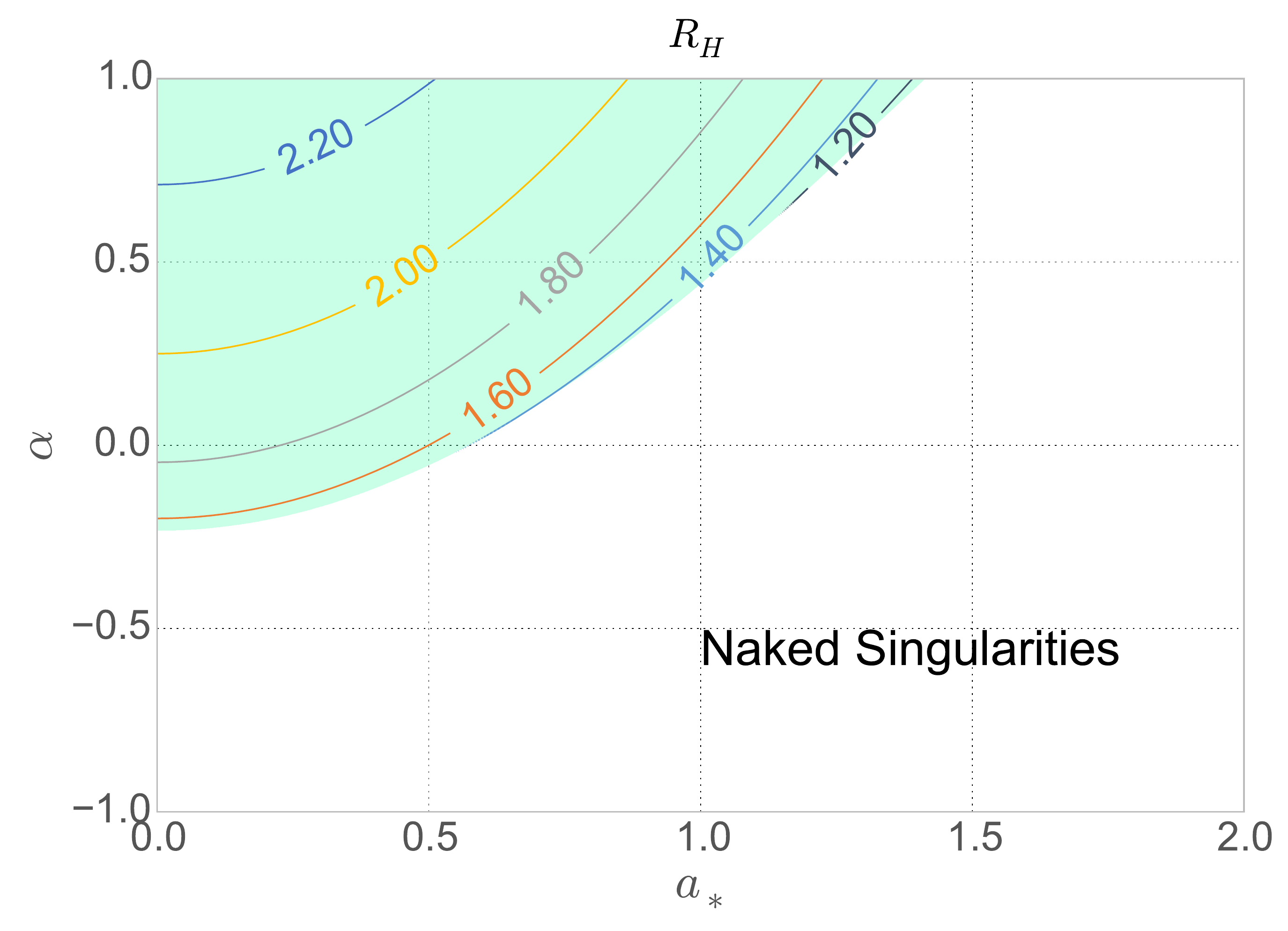}
\hspace{0.3cm}
\includegraphics[type=pdf,ext=.pdf,read=.pdf,width=8.5cm]{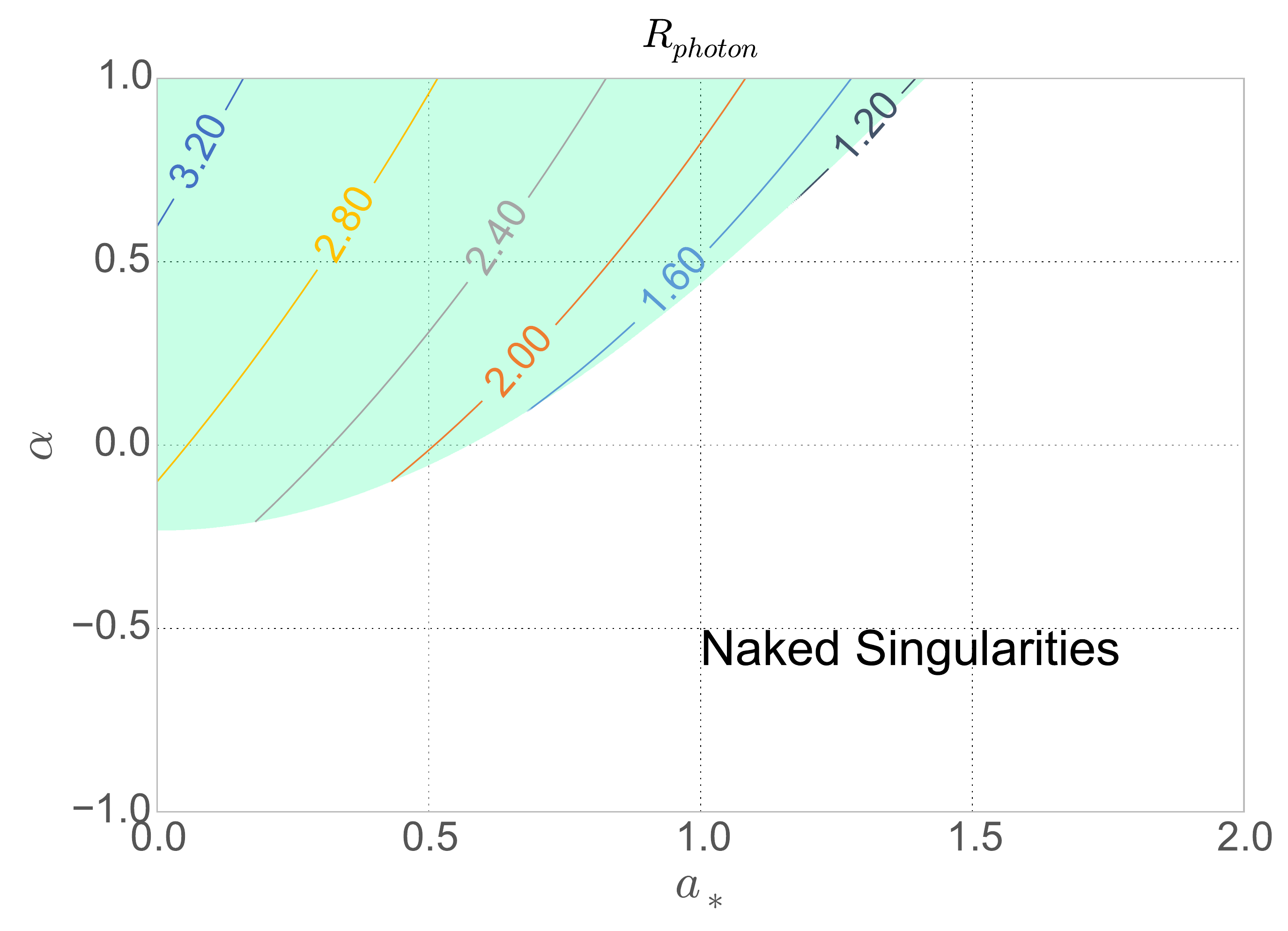} \\ 
\vspace{0.5cm}
\includegraphics[type=pdf,ext=.pdf,read=.pdf,width=8.5cm]{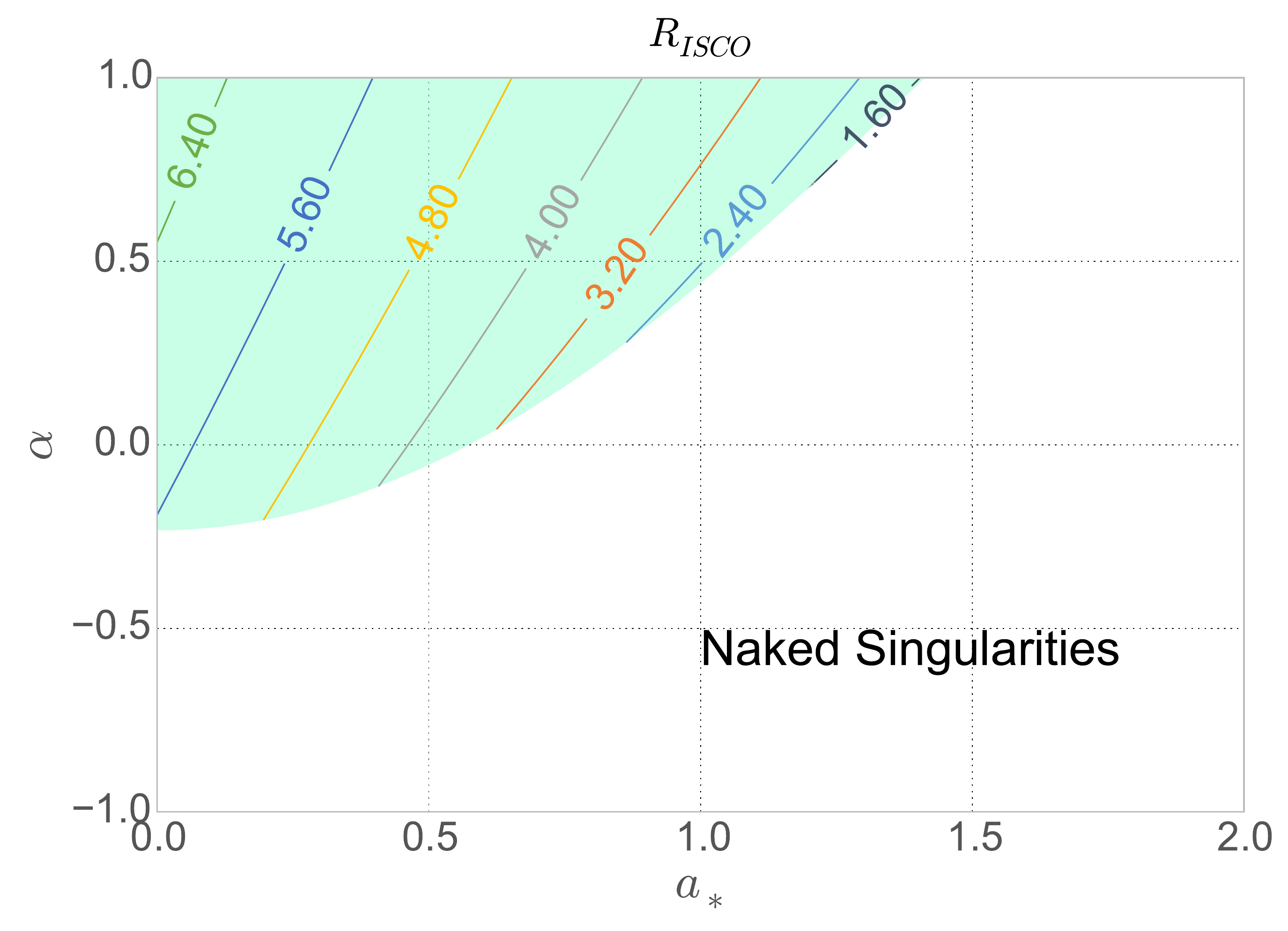}
\hspace{0.3cm}
\includegraphics[type=pdf,ext=.pdf,read=.pdf,width=8.5cm]{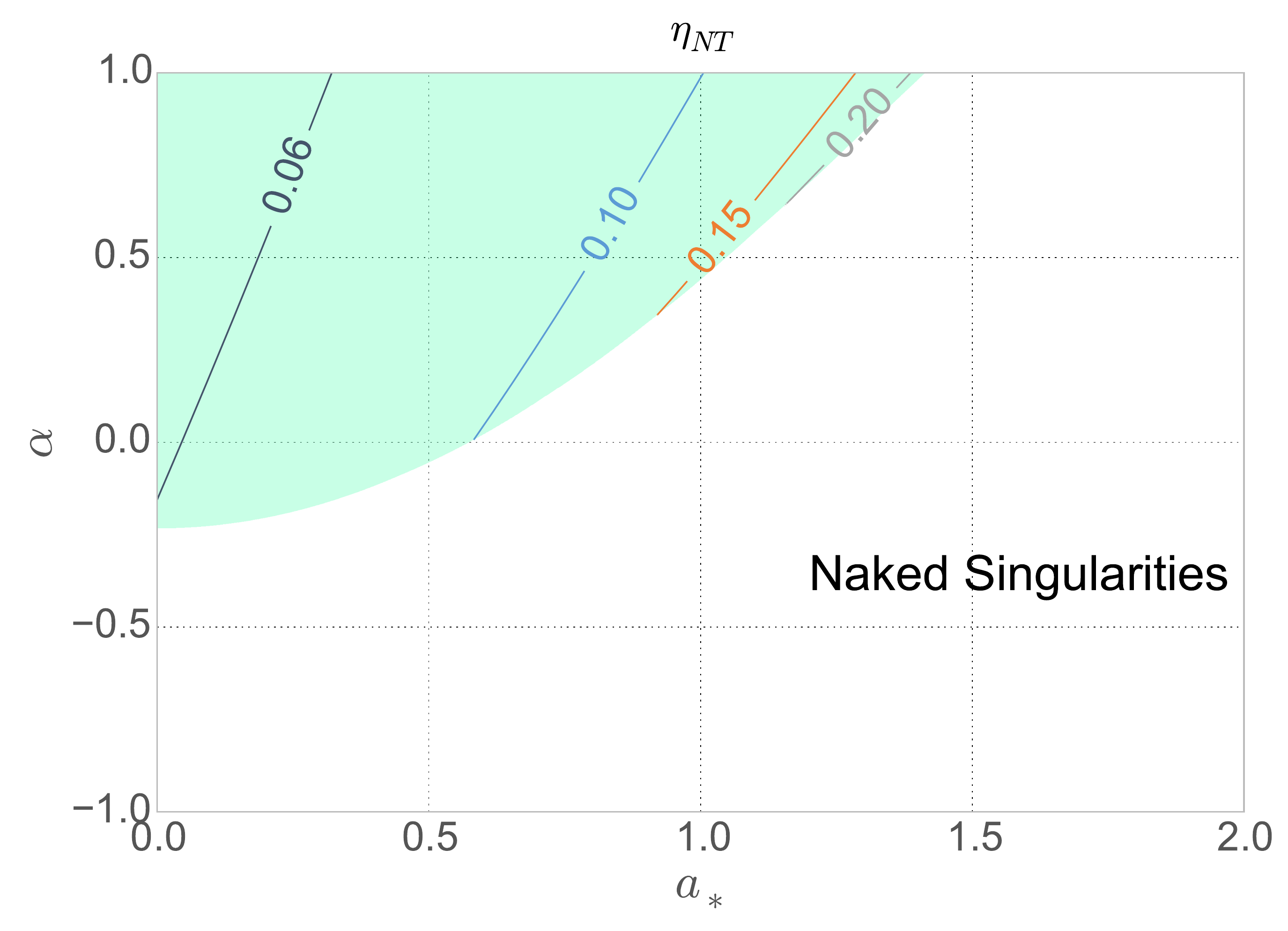}
\end{center}
\caption{As in Fig.~\ref{fig-b0.0} for $\beta = -0.5$.}
\label{fig-b-0.5}
\end{figure*}

\begin{figure*}
\begin{center}
\includegraphics[type=pdf,ext=.pdf,read=.pdf,width=8.5cm]{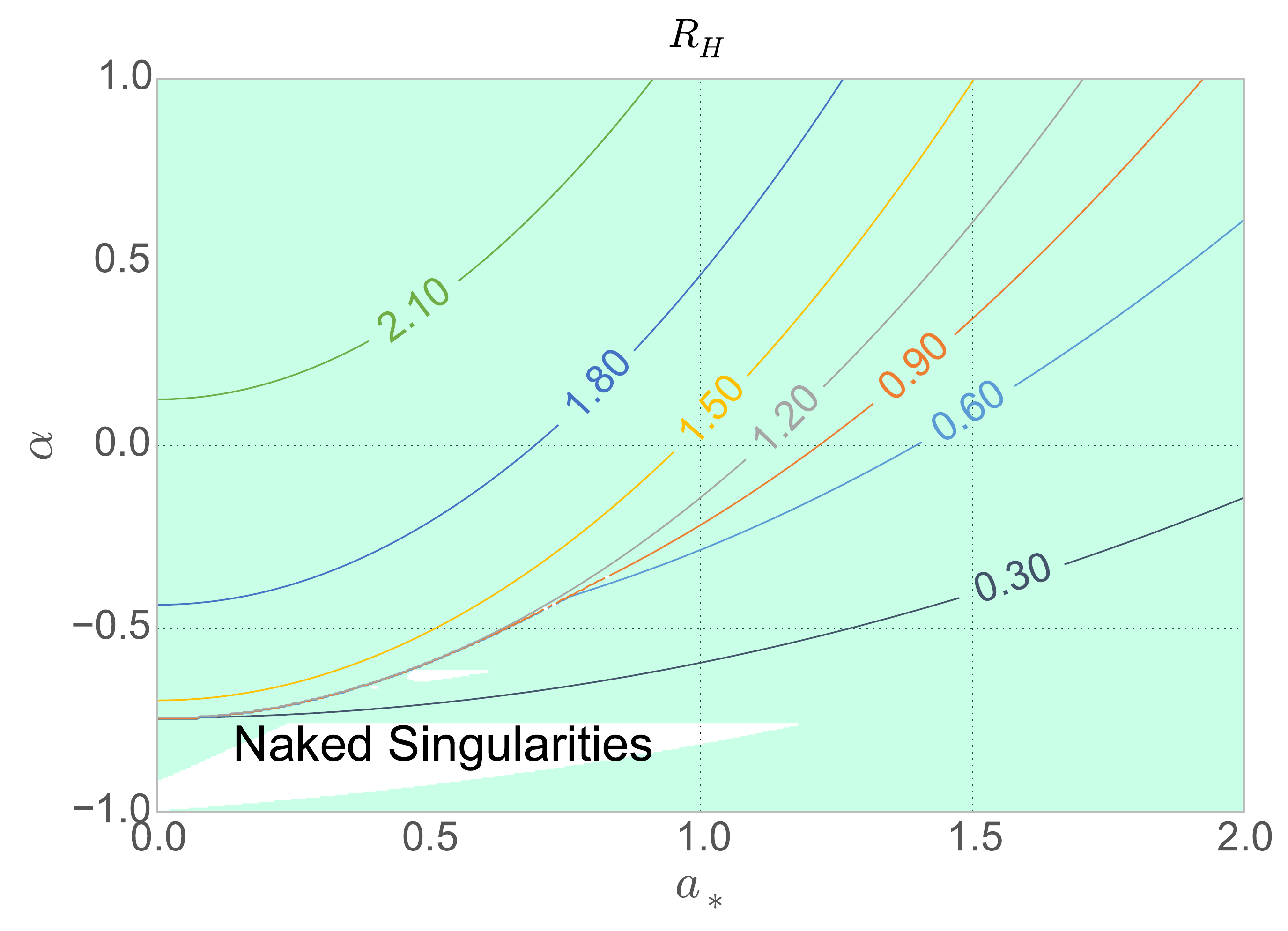}
\hspace{0.3cm}
\includegraphics[type=pdf,ext=.pdf,read=.pdf,width=8.5cm]{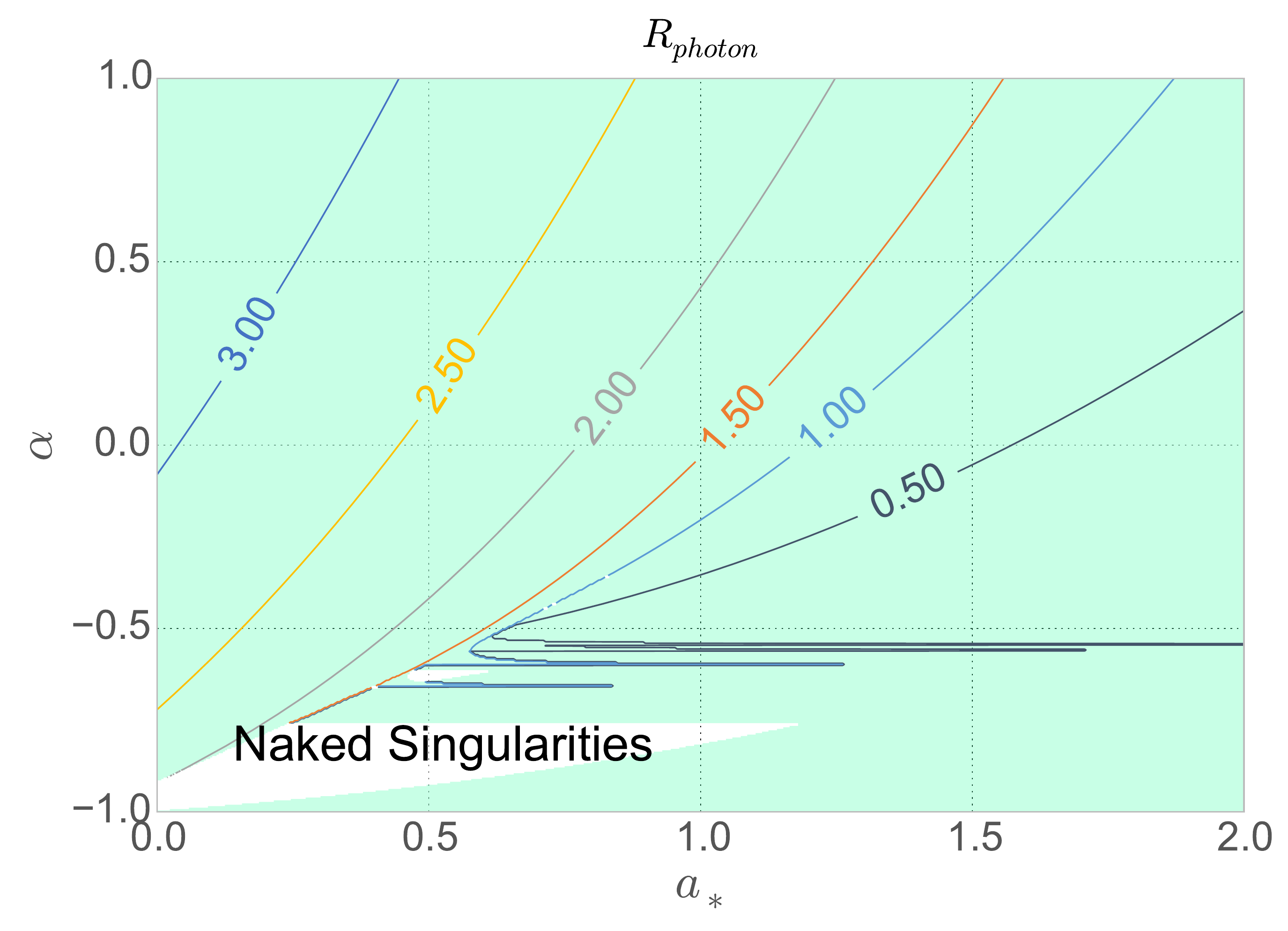} \\ 
\vspace{0.5cm}
\includegraphics[type=pdf,ext=.pdf,read=.pdf,width=8.5cm]{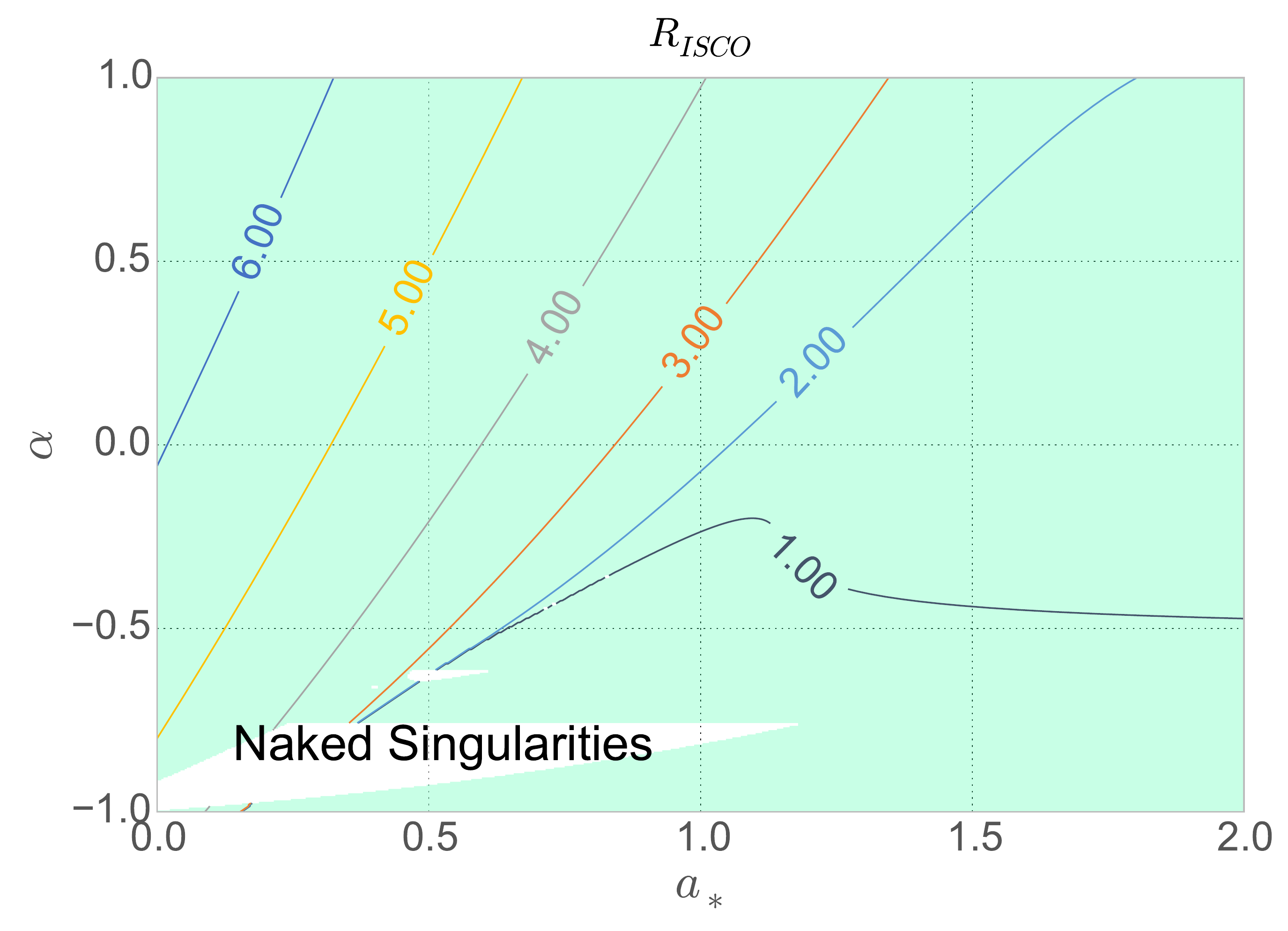}
\hspace{0.3cm}
\includegraphics[type=pdf,ext=.pdf,read=.pdf,width=8.5cm]{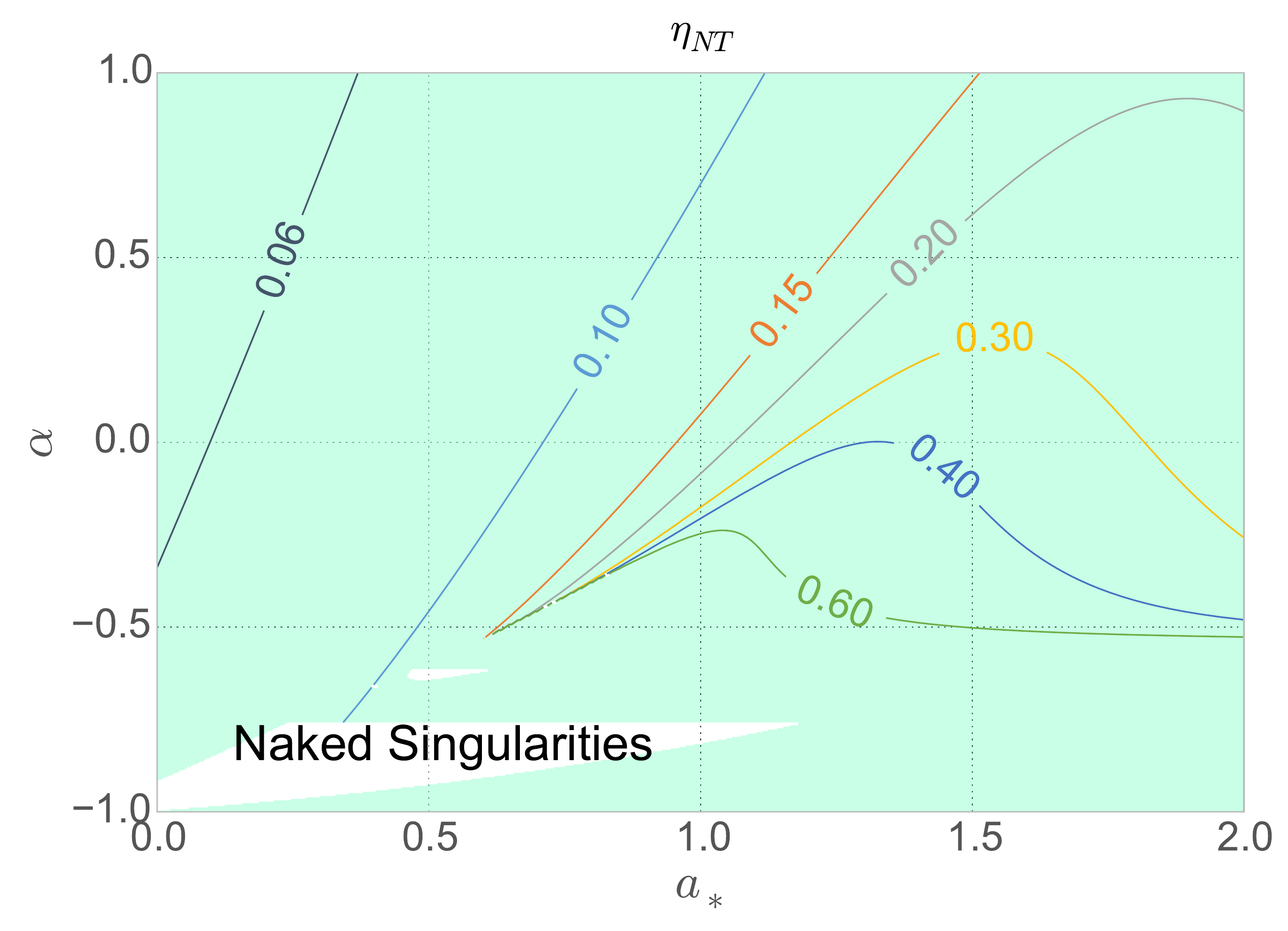}
\end{center}
\caption{As in Fig.~\ref{fig-b0.0} for $\beta = 0.2$.}
\label{fig-b0.2}
\end{figure*}

\begin{figure*}
\begin{center}
\includegraphics[type=pdf,ext=.pdf,read=.pdf,width=8.5cm]{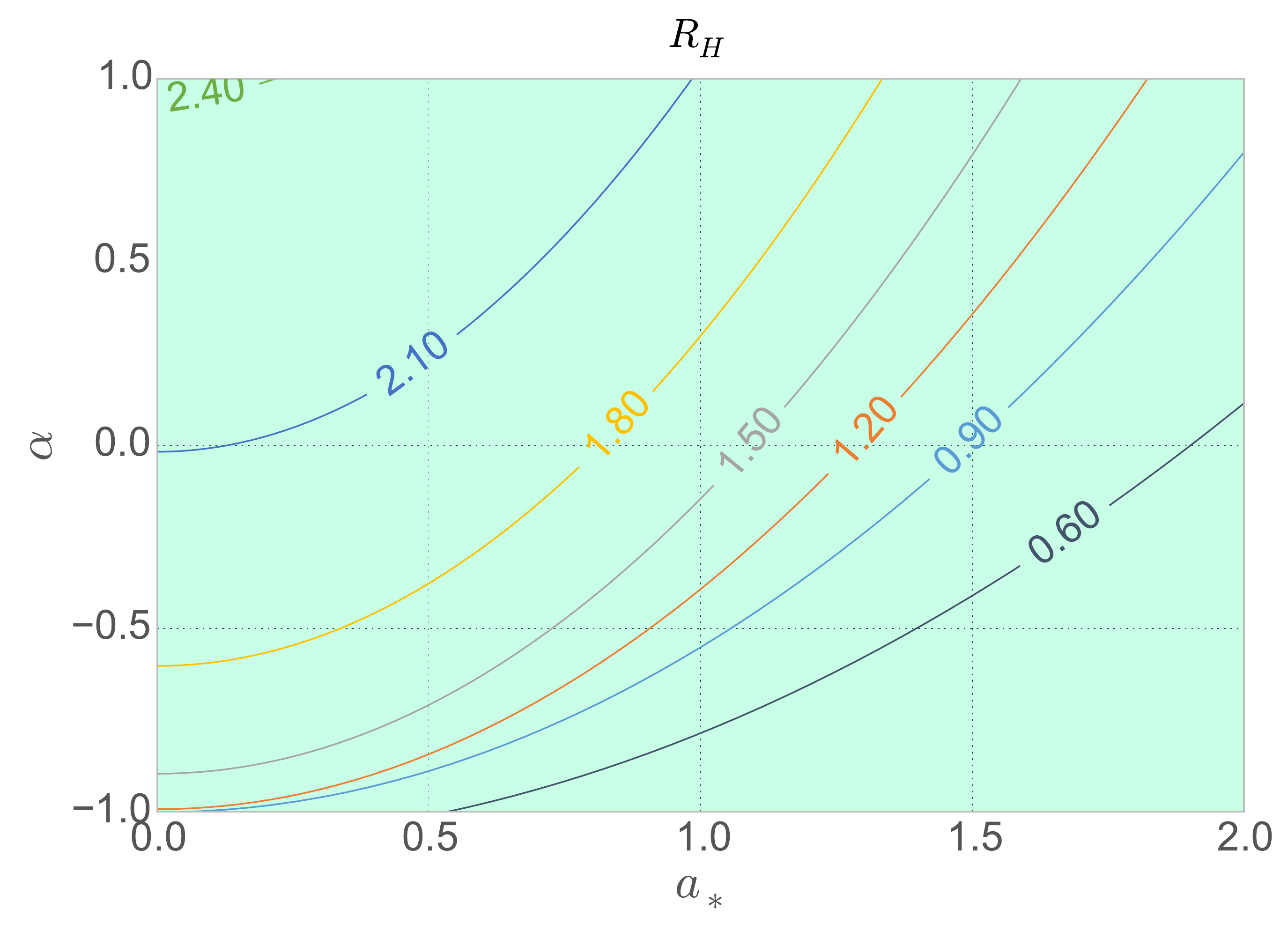}
\hspace{0.3cm}
\includegraphics[type=pdf,ext=.pdf,read=.pdf,width=8.5cm]{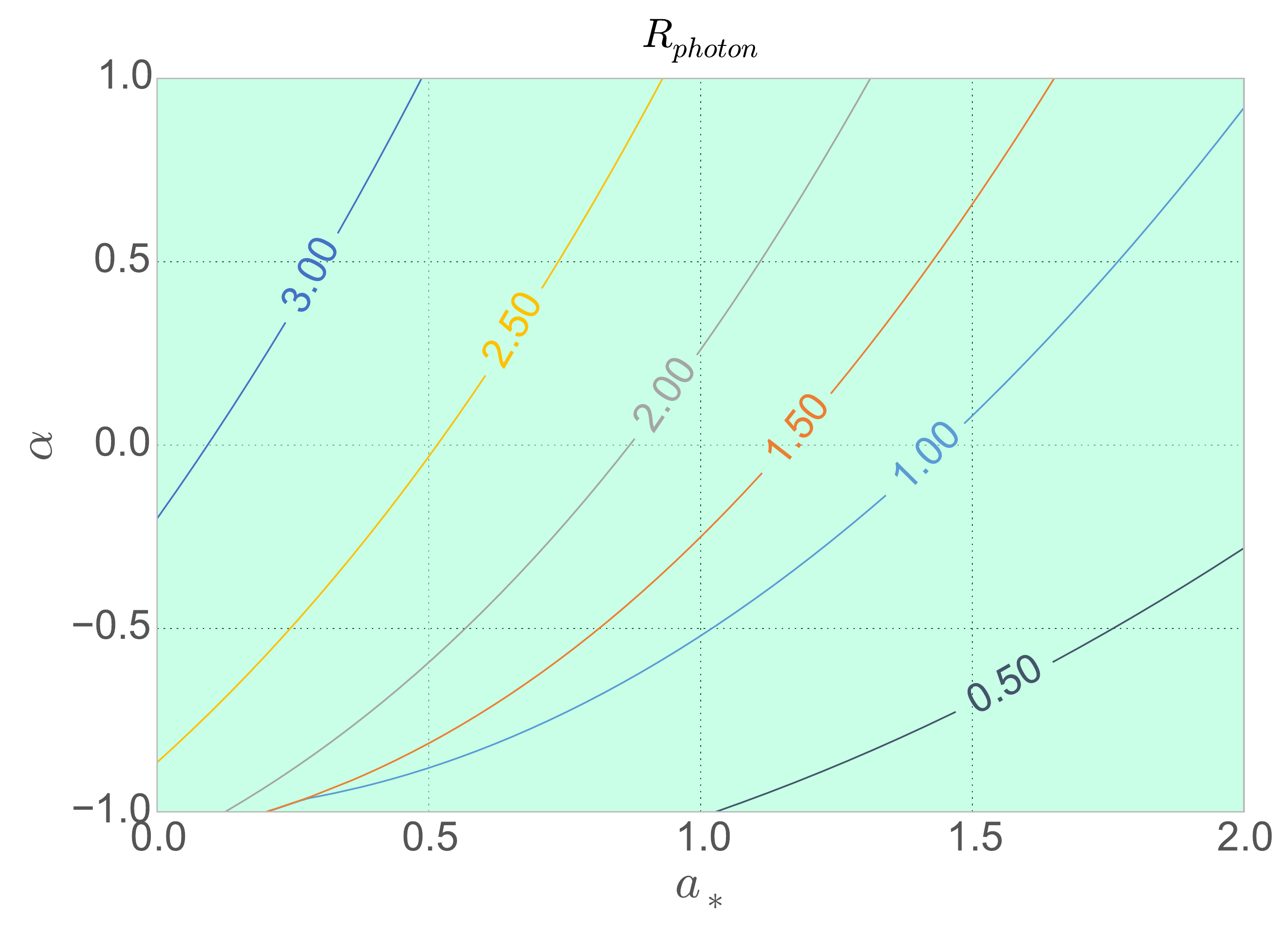} \\ 
\vspace{0.5cm}
\includegraphics[type=pdf,ext=.pdf,read=.pdf,width=8.5cm]{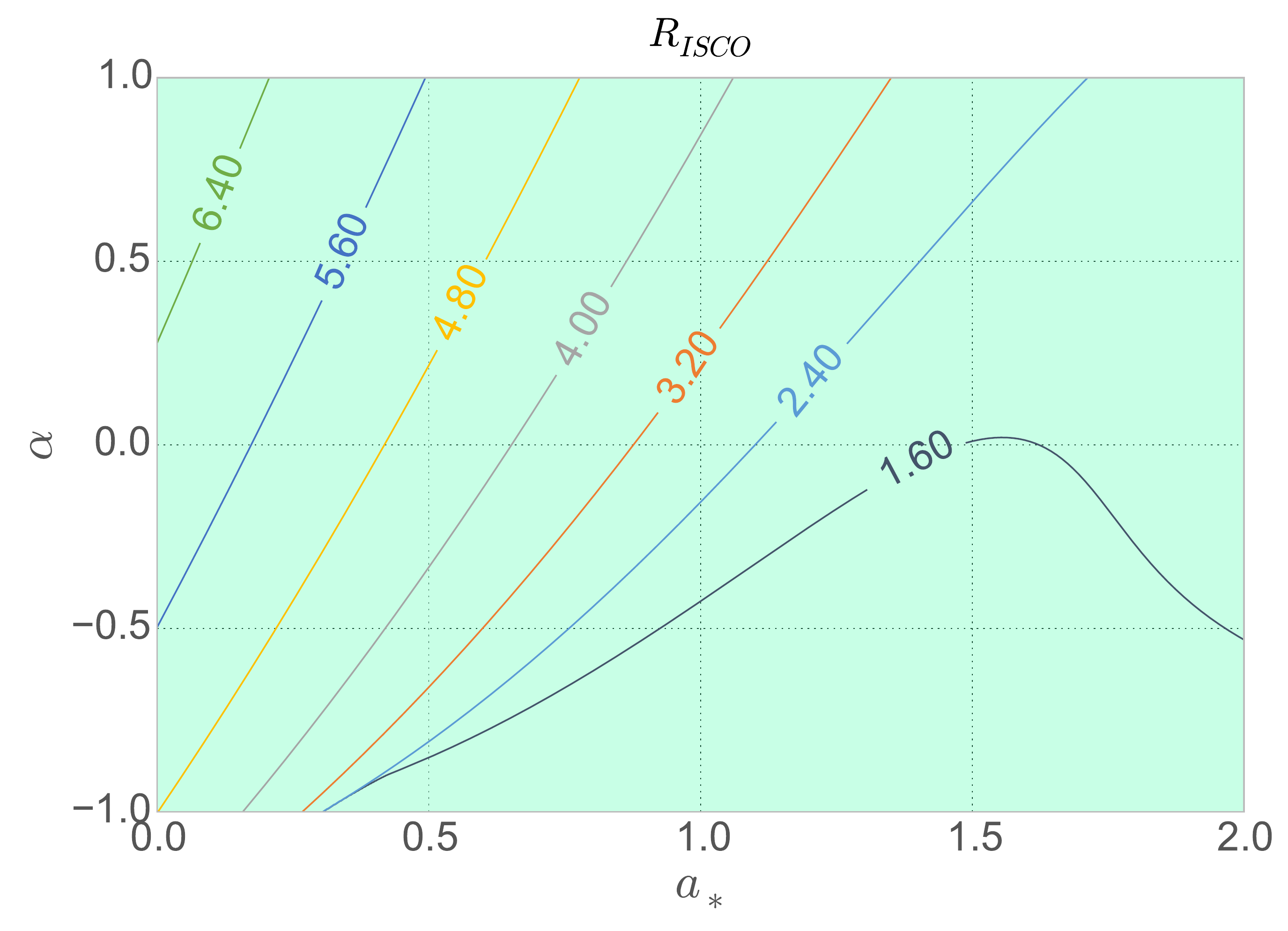}
\hspace{0.3cm}
\includegraphics[type=pdf,ext=.pdf,read=.pdf,width=8.5cm]{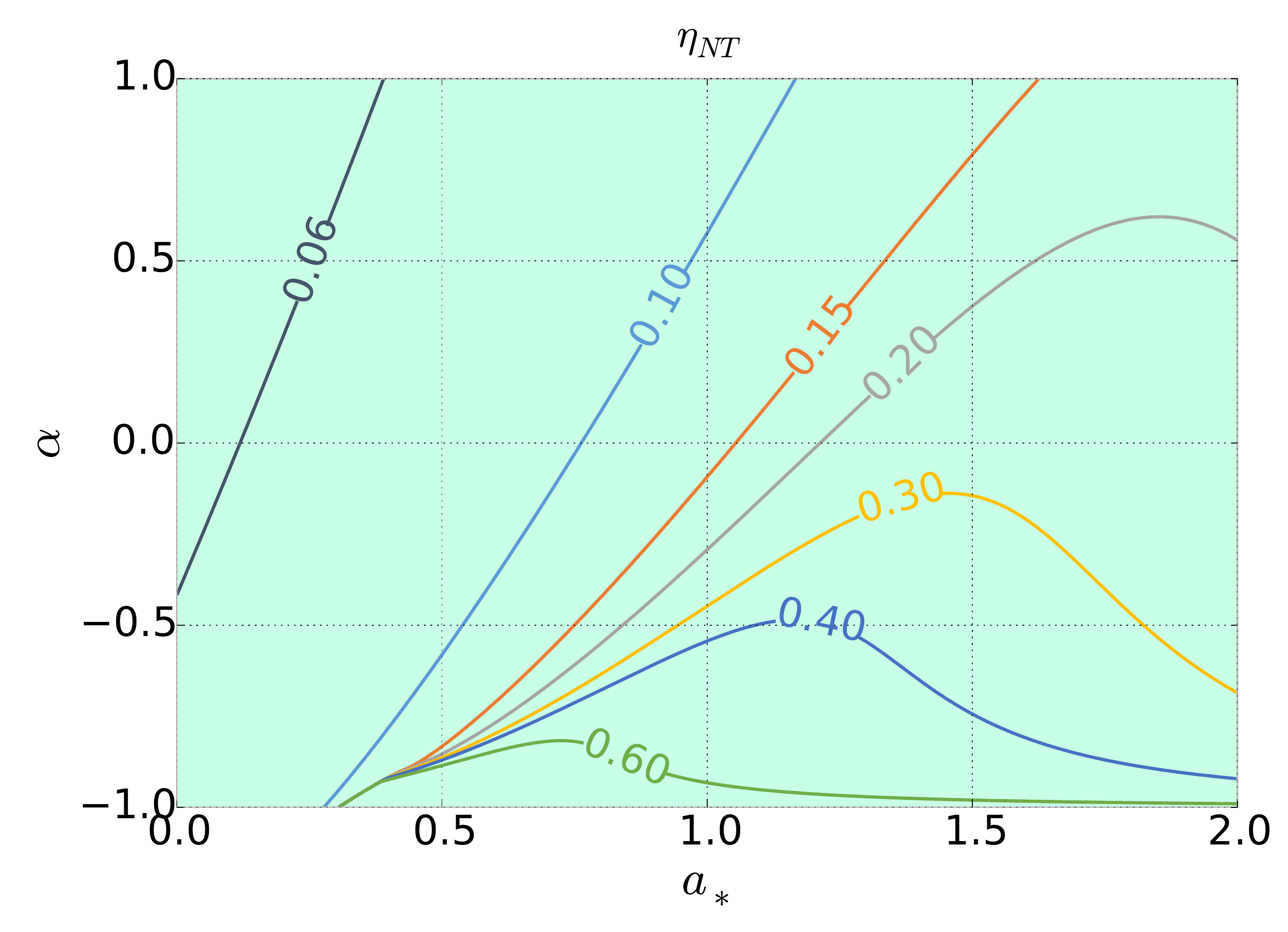}
\end{center}
\caption{As in Fig.~\ref{fig-b0.0} for $\beta = 0.5$.}
\label{fig-b0.5}
\end{figure*}

Figs.~\ref{fig-b0.0}-\ref{fig-b0.5} show the contour levels of the radius of the event horizon $R_{\rm H}$ (top left panels), of the photon radius $R_{\rm photon}$ (top right panels), of the ISCO radius $R_{\rm ISCO}$ (bottom left panels), and of the Novikov-Thorne radiative efficiency $\eta_{\rm NT}$ (bottom right panels) for $\beta = 0$ (Fig.~\ref{fig-b0.0}), $-0.2$ (Fig.~\ref{fig-b-0.2}), $-0.5$ (Fig.~\ref{fig-b-0.5}), $0.2$ (Fig.~\ref{fig-b0.2}), and $0.5$ (Fig.~\ref{fig-b0.5}).

The light-green regions are the parameter space with black holes, the white regions are those with naked singularities (no real solution of the equation $r^2 - 2 m_2 r + a^2 = 0$). $\alpha, \beta > 0$ make the gravitational force at small radii stronger (they ``increase'' the value of the effective mass), while $\alpha, \beta < 0$ make it weaker. The result is that for $\beta = 0$, $-0.2$, and $-0.5$, there may be naked singularity for $\alpha < 0$ because in those cases the gravitational force is not strong enough to create an event horizon. For $\beta = 0.5$, there are no naked singularities in the plots because at small radii the dominant term is $\beta M^3/r^3$ and it is positive (namely gravity is strong and there is an event horizon).

It is worth noting that the Novikov-Thorne radiative efficiency $\eta_{\rm NT}$ of our black holes may exceed the maximum value for a Kerr black hole $\eta_{\rm NT}^{\rm max} \approx 0.42$. This is not the case for most parametrizations proposed in the literature. If a black hole candidate has a high radiative efficiency, within the parametrizations in the literature it is quite automatic that the deviations from Kerr must be small, or otherwise it is impossible to reproduce the spectrum. In our case, this is not true, which means that our parametrization includes deviations from the Kerr solutions that are usually not taken into account. In a similar way, our black holes can be very small (the radius of the event horizon is small) and may also appear small (the photon capture radius is small). This is not the case in the other parametrizations.

\section{Hamilton-Jacobi equation}

The second ingredient necessary in the calculation of the spectrum of a thin disk is the determination of the photon trajectories from the point of emission to the point of detection. In a general spacetime, this is done by solving the geodesic equations, which are second order partial differential equations in the coordinates of the spacetime. The Kerr metric has the Carter constant and the equations of motion are separable and of first order. More importantly, the motion along the $\theta$ and $r$ directions can be reduced to elliptic integrals. The result is that numerical calculations can be faster and more accurate. In any non-trivial extension of the Kerr metric, this is not possible. However, we can have a metric in which the motion along the $\theta$ and $r$ directions can be reduced to hyper-elliptic integrals, with similar advantages of the Kerr solution.

The starting point is the Hamilton-Jacobi equation
\be
2 \frac{\partial S}{\partial \tau} = g^{\mu\nu}
\frac{\partial S}{\partial x^\mu} \frac{\partial S}{\partial x^\nu} \, .
\ee
Assuming in this section the more general case with $m_1$ and $m_2$ not necessarily the same, $g^{\mu\nu}$ is
\be
\left(\frac{\partial}{\partial s}\right)^2 &=&
- \frac{A_1}{\Sigma \Delta_1} \left(\frac{\partial}{\partial t}\right)^2
- \frac{4 a m_1 r}{\Sigma \Delta_1} 
\left(\frac{\partial}{\partial t}\right)\left(\frac{\partial}{\partial \phi}\right)
\nonumber\\
&&+ \frac{\Delta_2}{\Sigma} \left(\frac{\partial}{\partial r}\right)^2
+ \frac{1}{\Sigma} \left(\frac{\partial}{\partial \theta}\right)^2
\nonumber\\
&&+ \frac{\Delta_1 - a^2 \sin^2\theta}{\Sigma \Delta_1 \sin^2\theta} 
\left(\frac{\partial}{\partial \phi}\right)^2 \, ,
\ee
where $\Delta_1 = r^2 - 2 m_1 r + a^2$ and $A_1 = \left(r^2 + a^2\right)^2 - a^2 \Delta_1 \sin^2\theta$.

We can then proceed as in the Kerr case, looking for a solution of the Hamilton-Jacobi equation of the form (see Ref.~\cite{chandra}, Chapter~7, Section~62 for the details)
\be
S &=& - \frac{1}{2} \delta \tau - E t + L_z \phi S_r(r) + S_\theta(\theta) \, .
\ee
The solution for $S$ is
\be\label{eq-sssssss}
S &=& - \frac{1}{2} \delta \tau - E t + L_z \phi \nonumber\\ &&
+ \int^r \pm dr' \sqrt{\frac{R(r')}{\Delta_1 \Delta_2}}
+ \int^\theta \pm d\theta' \sqrt{\Theta(\theta')} \, ,
\ee
where $\delta = 1$ $(\delta = 0)$ for time-like (null) geodesics, $R(r)$ and $\Theta(\theta)$ are given by
\be
R(r) &=& \left[\left(r^2 + a^2\right) E - a L_z\right]^2 \nonumber\\ &&
- \Delta_1 \left[{\mathcal Q} + \left(L_z - a E\right)^2 + \delta r^2\right] \, , \\
\Theta(\theta) &=& {\mathcal Q} 
- \left[a^2 \left(\delta - E^2\right) + L_z^2 \csc^2\theta\right] \cos^2\theta \, ,
\ee
and the signs $\pm$ in~(\ref{eq-sssssss}) depend on the photon direction and they change at the turning points~\cite{chandra}. ${\mathcal Q}$ is the Carter constant, which reduces to the Carter constant of the Kerr metric when $\alpha=\beta=0$.

The equations of motion can be obtained by setting to zero the partial derivatives of $S$ with respect to the four constants of motion, $\delta$, $E$, $L_z$, and ${\mathcal Q}$. From $\partial S/\partial {\mathcal Q}=0$ we get
\be\label{eq-rt}
\int^r \pm dr' \sqrt{\frac{\Delta_1}{\Delta_2 R}} &=& \int^\theta \pm \frac{d\theta'}{\sqrt{\Theta}} \, . 
\ee
From $\partial S/\partial \delta=0$, $\partial S/\partial E=0$, and $\partial S/\partial L_z=0$, we find, respectively,
\be
\tau &=& \int^r dr' r'^2 \sqrt{\frac{\Delta_1}{\Delta_2 R}} 
+ a^2 \int^\theta d\theta' \frac{\cos^2\theta'}{\sqrt{\Theta}}\, , \\
t &=& \tau E \nonumber\\ && + 2 
\int^r  \frac{m_1(r') dr'}{\sqrt{\Delta_1 \Delta_2 R}} \left[r'^2 E - a \left(L_z - a E\right)\right] r' \, , \\
\phi &=& a \int^r \frac{dr'}{\sqrt{\Delta_1 \Delta_2 R}} \left[\left(r'^2 + a^2\right) E - a L_z\right] 
\nonumber\\
&& + \int^\theta \frac{d\theta'}{\sqrt{\Theta}} \left(L_z \csc^2\theta' - a E\right)\, .
\label{eq-phi}
\ee
As in the Kerr metric~\cite{chandra}, it is straightforward to verify that the system of equations (\ref{eq-rt})-(\ref{eq-phi}) is equivalent to the set of equations 
\be
\Sigma^2 \dot{r}^2 &=& \frac{\Delta_2}{\Delta_1} R \, , \\
\Sigma^2 \dot{\theta}^2 &=& \Theta \, , \\
\Sigma \dot{\phi} &=& \frac{1}{\Delta_1} \left[2 a m_1 r E + \left(\Sigma - 2 m_1 r\right) L_z \csc^2\theta\right] \, , \\
\Sigma \dot{t} &=& \frac{1}{\Delta_1} \left(A_1 E - 2 a m_1 r L_z\right) \, .
\ee

\section{Motion of massless particles from the disk to the observer \label{s-sss}}

We want now to study the motion of the photons from the point of emission in the disk to the point of detection at infinity. We assume the choice in Eq.~(\ref{eq-Mab}) and we define $m = m_1 = m_2$. For null geodesics, $\delta=0$ and we use the parameters $\xi = -k_\phi/k_t$ and $q^2 = {\mathcal Q}/k_t^2$, where $k_t=-E$ and $k_\phi = L_z$. This choice is useful for $\delta=0$ because the photon trajectories are independent of the photon energy. We have
\be
\tilde{R} &=& \frac{R}{E^2} = r^4 + \left(a^2 - \xi^2 - q^2\right) r^2 \nonumber\\ && \hspace{1cm}
+ 2 m \left[q^2 + \left(\xi - a\right)^2\right] r - a^2 q^2 \, , \\
\tilde{\Theta} &=& \frac{\Theta}{E^2} = q^2 + a^2 \cos^2\theta - \xi^2 \cot^2\theta \, .
\ee
The relation between the parameters $(\xi,q)$ and the celestial coordinates $(X,Y)$ of the image as seen by an observer at infinity is the same as in Kerr (because it requires $r\rar \infty$)~\cite{chandra}
\be
X &=& \lim_{r \rar \infty} \left(\frac{r p^{(\phi)}}{p^{(t)}}\right) = \xi \csc\theta_{\rm o} \, , \\
Y &=& \lim_{r \rar \infty} \left(\frac{r p^{(\theta)}}{p^{(t)}}\right) \nonumber\\ &=& 
\pm \sqrt{q^2 + a^2 \cos^2\theta_{\rm o} - \xi^2 \cot^2\theta_{\rm o}} \, ,
\ee
where $\theta_{\rm o}$ is the angular coordinate of the observer at infinity.

In the calculation of the photons from the disk to the distant observer, we are only interested in the motion on the $(r,\theta)$-plane. The master equation is
\be\label{eq-rrrr}
\int^{r_{\rm o}}_{r_{\rm e}} \frac{dr'}{\sqrt{\tilde{R}}} 
&=& \int^{\theta_{\rm o}}_{\pi/2} \frac{d\theta'}{\sqrt{\tilde{\Theta}}} \, , 
\ee
where $r_{\rm o}$ is the radial coordinate of the observer at infinity, $r_{\rm e}$ is the radius of the emission point on the disk, and $\theta_{\rm e} = \pi/2$ because the disk is in the equatorial plane.

Since $\tilde{\Theta}$ is independent of $\alpha$ and $\beta$, one can use the same calculation technique as in Kerr~\cite{chandra,c75,lixin,naoc}. The integral can be transformed to
\be
\int^{\theta_{\rm o}}_{\pi/2} \frac{d\theta'}{\sqrt{\tilde{\Theta}}} 
= C_\theta F[\psi_\theta (\pi/2), \kappa_\theta] \, ,
\ee
where $F$ is an elliptic integral of the first kind with argument $\psi_\theta$ and modulus $\kappa_\theta$, while $C_\theta$, $\psi_\theta$, and $\kappa_\theta$ are functions of $\xi$ and $q$.

In the Kerr case, even the integral in $r$ can be reduced to an elliptic integral of the first kind. This is not possible in a non-trivial generalization of the Kerr metric, because in the Kerr metric the function $\tilde{R}$ in Eq.~(\ref{eq-rrrr}) is already a polynomial of fourth order. However, the integral can be transformed to a hyper-elliptic integral, see Ref.~\cite{2011}. The calculations are somewhat more difficult, but the procedure is well known. In the case of our default choice~(\ref{eq-Mab}), we have
\be\label{eq-int-r}
\int^{r_{\rm o}}_{r_{\rm e}} \frac{dr'}{\sqrt{\tilde{R}}} &=& \int^{r_{\rm o}}_{r_{\rm e}} \frac{r' \, dr'}{\sqrt{P_6(r')}}
\ee
where $P_6(r)$ is a polynomial of order 6
\be
P_6(r) &=& r^6 + \left(a^2 - \xi^2 - \eta\right) r^4 
+ 2 M \left[q^2 + \left(\xi - a\right)^2\right] r^3 
\nonumber\\ &&
- a^2 q^2 r^2
+ 2 \alpha M^3 \left[q^2 + \left(\xi - a\right)^2\right] r
\nonumber\\ &&
+ 2 \beta M^4 \left[q^2 + \left(\xi - a\right)^2\right] \, .
\ee
In the end, it is possible to write Eq.~(\ref{eq-int-r}) as a function of $\xi$, $q$, and $r_{\rm e}$ and solve Eq.~(\ref{eq-rrrr}) in terms of $r_{\rm e}$:
\be\label{eq-r-r}
r_{\rm e} = r_{\rm e}(\xi,q)
\ee

\section{Transfer function}

The calculation of the spectrum of thin disks can be conveniently split into two parts: one concerning the background metric, and another part related to the astrophysical model. In this way, we can focus our attention on the relativistic effects determined by the background metric, which can be later combined with the astrophysical models already discussed in the literature. This can be done by introducing the transfer function $f$, which takes into account all the relativistic effects (gravitational redshift, Doppler boosting, light bending)~\cite{c75,srr}.

The observed flux is
\be
F_{\rm o} (\nu_{\rm o}) 
= \int I_{\rm o}(\nu_{\rm o}) d\tilde{\Omega} 
= \int g^3 I_{\rm e}(\nu_{\rm e}) d\tilde{\Omega} \, ,
\ee
where $I_{\rm o}$ and $I_{\rm e}$ are, respectively, the specific intensities of the radiation detected by the distant observer and the specific intensities of the radiation as measured by the emitter, 
$d\tilde{\Omega} = dX dY/r^2_{\rm o}$ is the element of the solid angle subtended by the image of the disk on the observer's sky, $r_{\rm o}$ is the distance of the observer from the source, and $I_{\rm o} = g^3 I_{\rm e}$ follows from Liouville's theorem. $I_{\rm e}$ is a function of the emitted frequency $\nu_{\rm e}$, but also of the emission radius $r_{\rm e}$ and of the direction of the photon emission, which can be described by the polar angle $n_{\rm e}$ of the emitted photon with respect to the normal of the disk in the rest frame of the gas.

It is convenient to introduce the relative redshift $g^* = g^* (r_{\rm e},\theta_{\rm o})$, defined by 
\be
g^* = \frac{g - g_{\rm min}}{g_{\rm max} - g_{\rm min}} \, ,
\ee
which ranges from 0 to 1. Here $g_{\rm max}=g_{\rm max}(r_{\rm e},\theta_{\rm o})$ and $g_{\rm min}=g_{\rm min}(r_{\rm e},\theta_{\rm o})$ are, respectively, the maximum and the minimum values of $g$ for the photons emitted from the radial coordinate $r_{\rm e}$ and detected by a distant observer with polar coordinate $\theta_{\rm o}$. The observed flux can now be rewritten as
\begin{widetext}
\be
F_{\rm o} (\nu_{\rm o}) 
= \frac{1}{r^2_{\rm o}} \int_{R_{\rm ISCO}}^{\infty} \int_0^1
\pi r_{\rm e} \frac{ g^2}{\sqrt{g^* (1 - g^*)}} f(g^*,r_{\rm e},\theta_{\rm o})
I_{\rm e}(\nu_{\rm e},r_{\rm e},n_{\rm e}) \, dg^* \, dr_{\rm e}
\ee
\end{widetext}
where $f$ is the transfer function and takes into account all the relativistic effects determined by the background metric
\be
f(g^*,r_{\rm e},\theta_{\rm o}) = \frac{1}{\pi r_{\rm e}} g 
\sqrt{g^* (1 - g^*)} \left| \frac{\partial \left(X,Y\right)}{\partial \left(g^*,r_{\rm e}\right)} \right| \, .
\ee
Since
\be
\left| \frac{\partial \left(X,Y\right)}{\partial \left(g^*,r_{\rm e}\right)} \right|
= \frac{q \left(g_{\rm max} - g_{\rm min}\right)}{Y \sin\theta_{\rm o}}
\left| \frac{\partial \left(\xi,q\right)}{\partial \left(g,r_{\rm e}\right)} \right| \, ,
\ee
the calculation of the transfer function $f$ requires the evaluation of the Jacobian
\be
\left| \frac{\partial \left(\xi,q\right)}{\partial \left(g,r_{\rm e}\right)} \right| = 
\left| \frac{\partial \xi}{\partial g} \frac{\partial q}{\partial r_{\rm e}} -
\frac{\partial q}{\partial g} \frac{\partial \xi}{\partial r_{\rm e}}\right| \, ,
\ee
which can be done numerically from Eqs.~(\ref{eq-g-g}) and (\ref{eq-r-r}) in the same way as in the Kerr case~\cite{srr}.

Lastly, if the transfer function depends on the emission angle $n_{\rm e}$, this is given by~\cite{srr}
\be
\cos(n_{\rm e}) = - \frac{n^\mu k_\mu}{u^\nu_{\rm e} k_\nu} = \frac{q g}{r_{\rm e}} \, ,
\ee
because $n^\mu = \left(0, 0, 1/r_{\rm e}, 0\right)$, $u^\nu_{\rm e} k_\nu = k_t/g$, and $q = - k_\theta/k_t$.

\begin{figure*}[t]
\begin{center}
\includegraphics[type=pdf,ext=.pdf,read=.pdf,width=8.5cm]{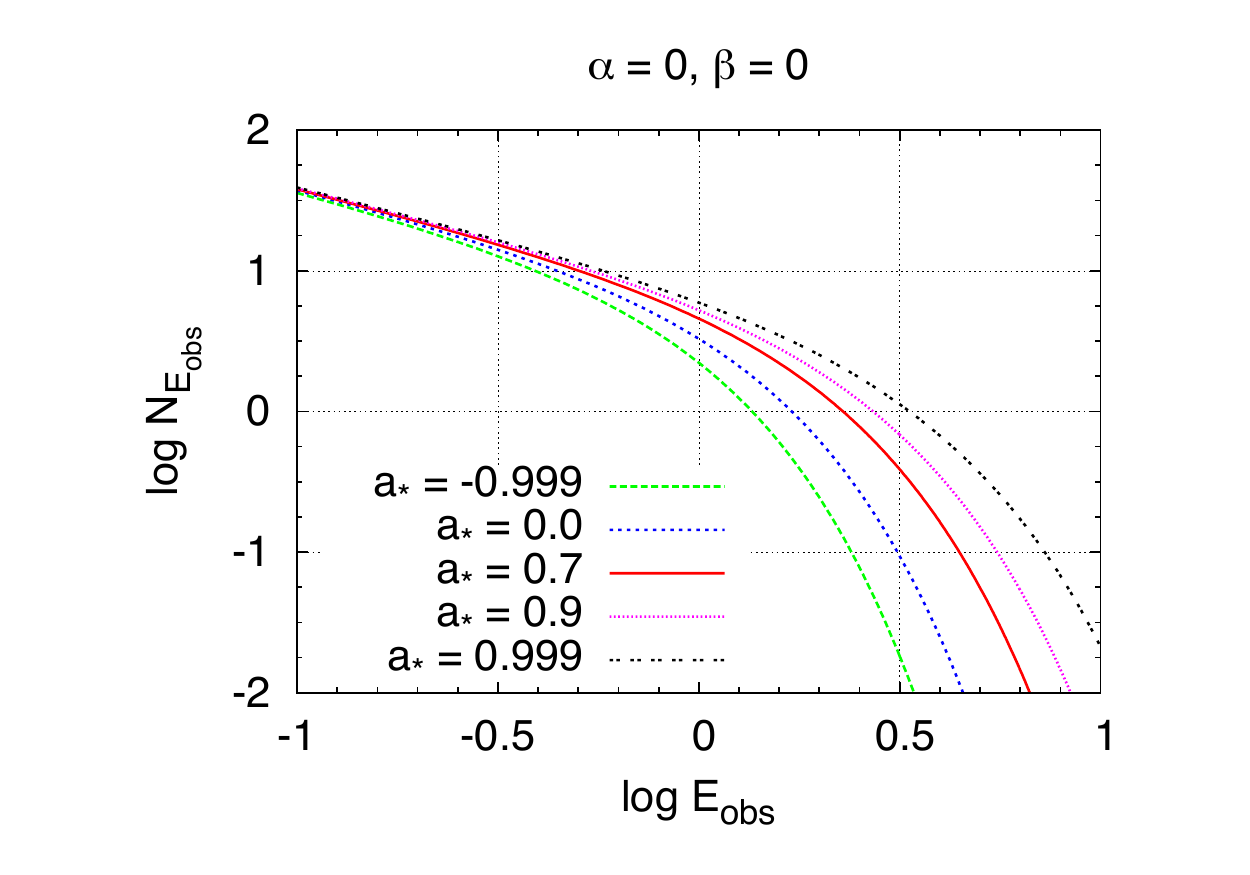}
\hspace{-0.3cm}
\includegraphics[type=pdf,ext=.pdf,read=.pdf,width=8.5cm]{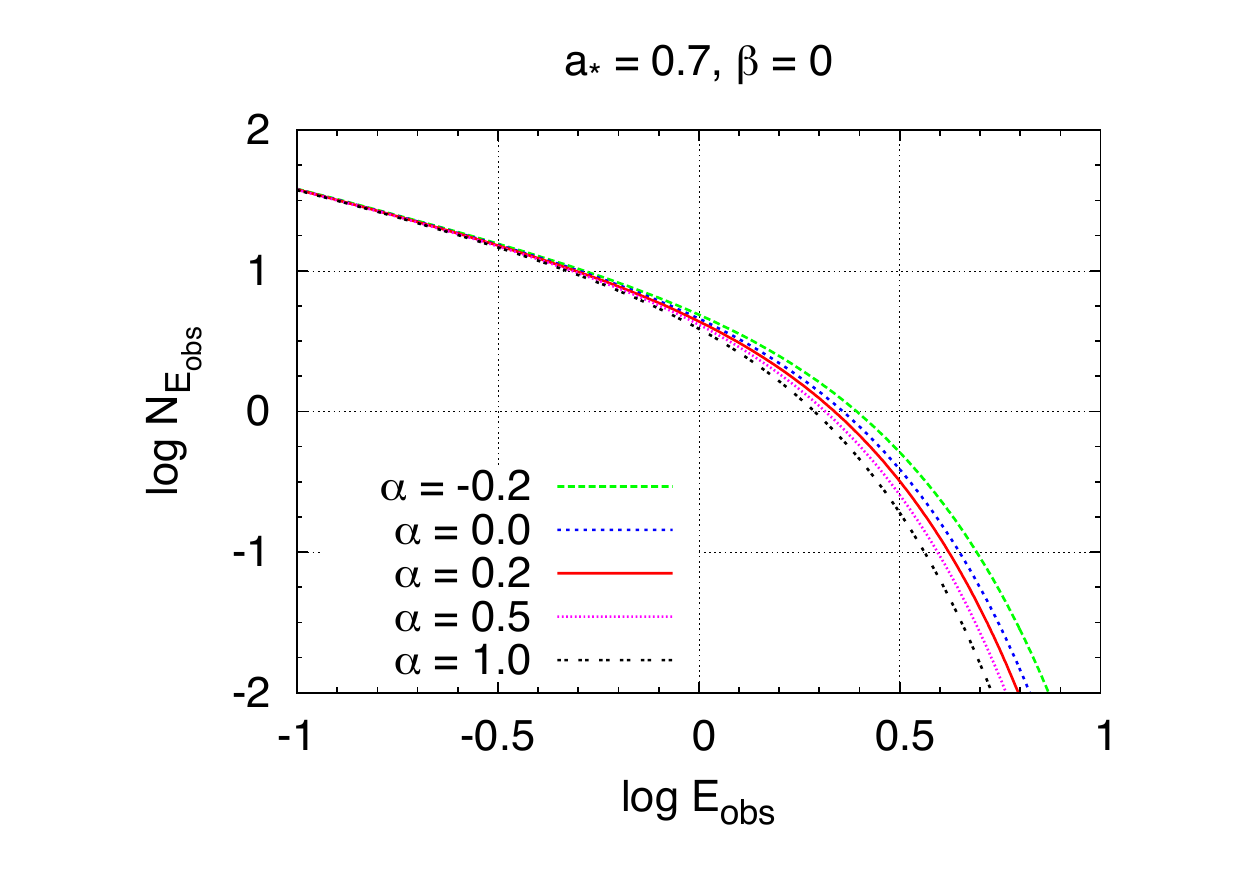} \\ 
\vspace{0.2cm}
\includegraphics[type=pdf,ext=.pdf,read=.pdf,width=8.5cm]{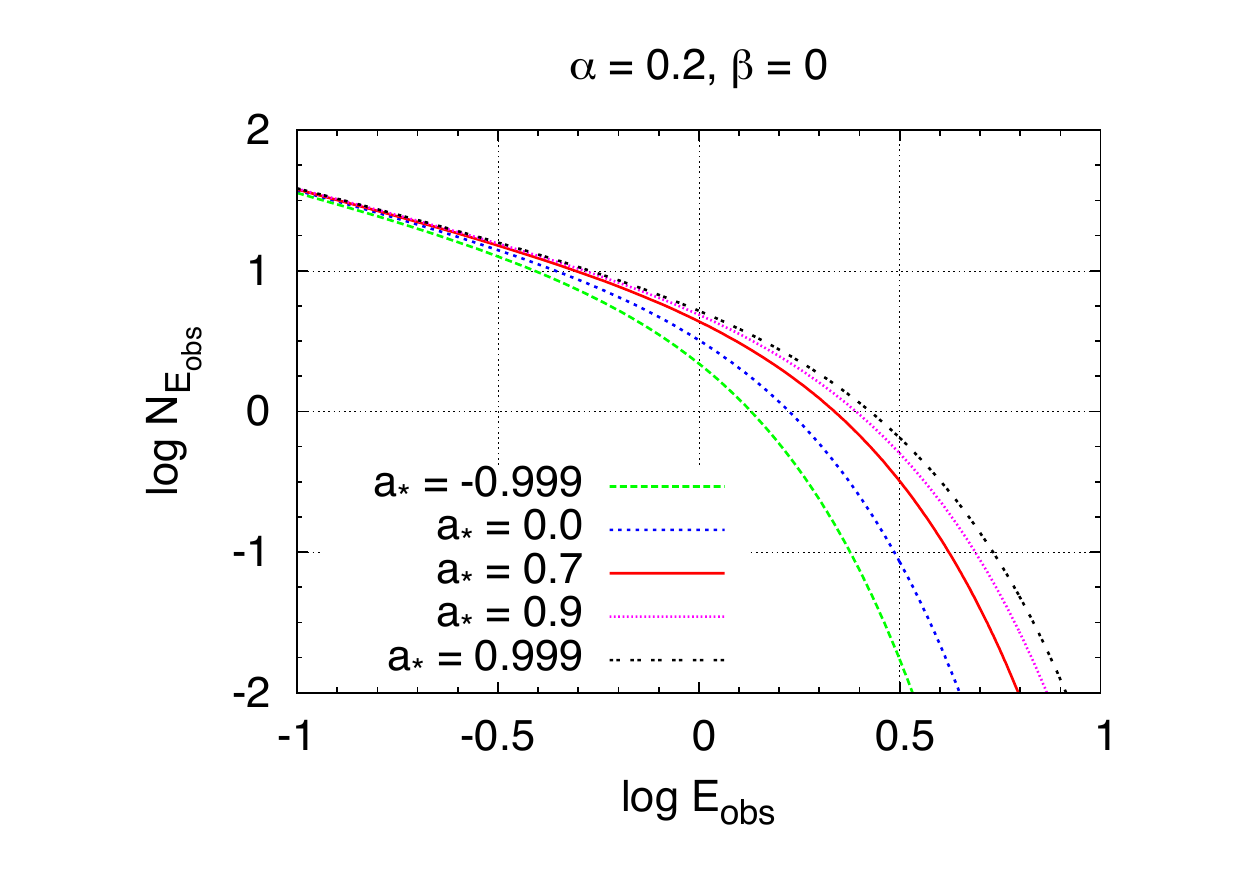}
\hspace{-0.3cm}
\includegraphics[type=pdf,ext=.pdf,read=.pdf,width=8.5cm]{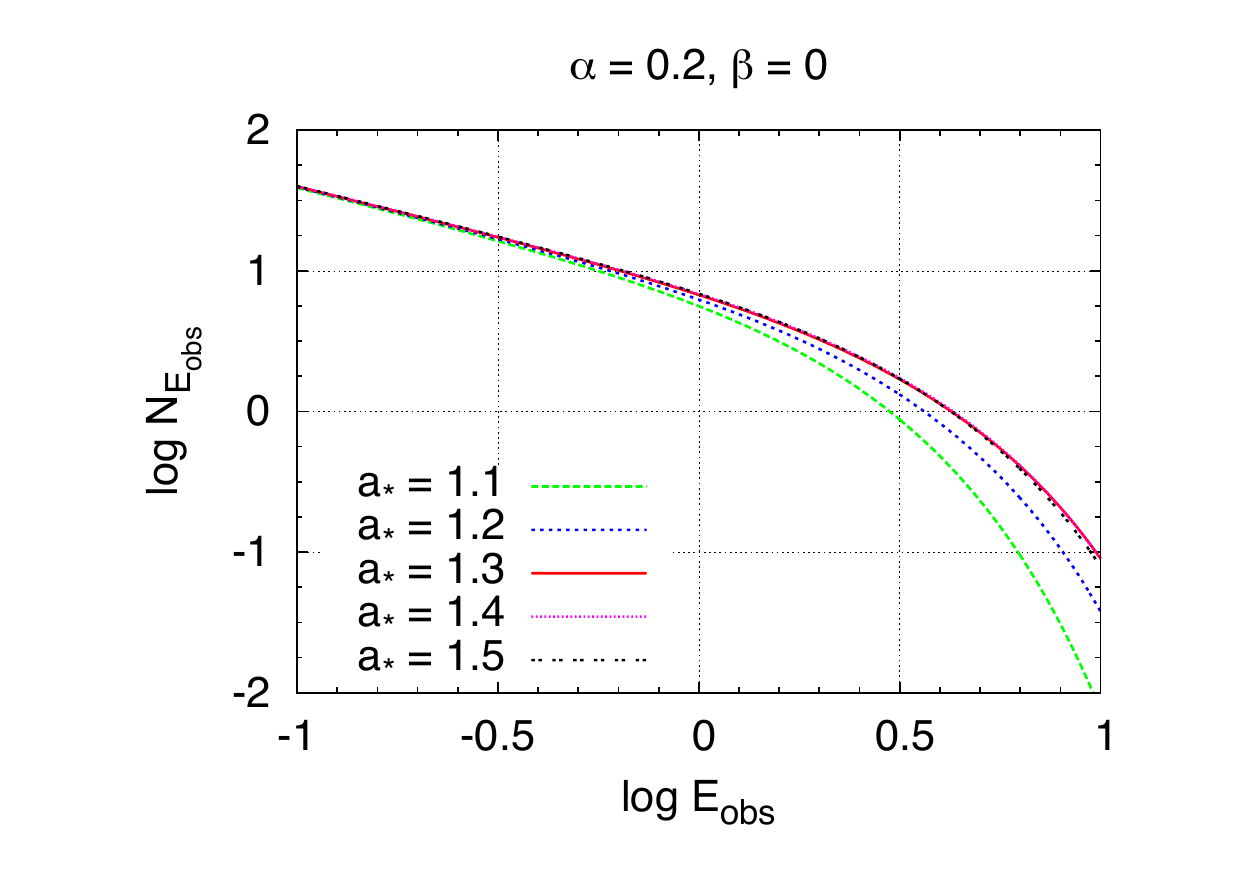}
\end{center}
\vspace{-0.3cm}
\caption{Impact of the deformation parameter $\alpha$ on the thermal spectrum of thin disks. Top left panel: spectra in Kerr spacetimes with different values of the spin parameter. Top right panel: spectra in spacetimes with spin parameter $a_* = 0.7$ and different values of the deformation parameter $\alpha$ ($\beta = 0$). Bottom panels: spectra in non-Kerr spacetimes with $\alpha = 0.2$ and different values of the spin parameter $a_*$. The values of the other model parameters are: mass $M = 10$~$M_\odot$, mass accretion rate $\dot{M} = 2 \cdot 10^{18}$~g~s$^{-1}$, distance $D = 10$~kpc, viewing angle $i = 45^\circ$, color factor $f_{\rm col}=1.6$, and $\Upsilon = 1$. Flux density $N_{E_{\rm obs}}$ in photons~keV$^{-1}$~cm$^{-2}$~s$^{-1}$, and photon energy $E_{\rm obs}$ in keV. See the text for more details.}
\label{fig-cfm1}
\end{figure*}

\section{Thermal spectrum of thin disks}

While it is not the purpose of this paper to study the observational implications of our metric and to constrain $\alpha$ and $\beta$, it is useful to understand the impact of these deformation parameters on the spectrum of a black hole. To do this, we consider the thermal spectrum of a thin disk. In this case, the specific intensities of the radiation in the rest-frame of the gas is (for more details, see e.g.~\cite{tt2} and reference therein)
\be\label{eq-i-bb}
I_{\rm e}(\nu_{\rm e},r_{\rm e},n_{\rm e}) = \frac{2 h \nu^3_{\rm e}}{c^2} \frac{1}{f_{\rm col}^4} 
\frac{\Upsilon(n_{\rm e})}{\exp\left(\frac{h \nu_{\rm e}}{k_{\rm B} T_{\rm col}(r)}\right) - 1} \, ,
\ee
where $h$ is the Planck's constant, $c$ is the speed of light, $k_{\rm B}$ is the Boltzmann constant, $\Upsilon(n_{\rm e})$ is a function that depends on the emission model (for example, $\Upsilon = 1$ for isotropic emission and $\Upsilon = \frac{1}{2} + \frac{3}{4} n_{\rm e}$ for limb-darkened emission), and $f_{\rm col} \approx 1.6$ is the color (or hardening) factor. The color temperature is $T_{\rm col}(r) = f_{\rm col} T_{\rm eff} (r)$, where $T_{\rm eff} (r)$ is the effective temperature in the Novikov-Thorne model defined as $\mathcal{F}(r) = \sigma T^4_{\rm eff}$. $\sigma$ is the Stefan-Boltzmann constant and $\mathcal{F}(r)$ is the time-averaged energy flux from the surface of the disk
\be
\hspace{-0.2cm}
\mathcal{F}(r) = \frac{\dot{M}}{4 \pi \sqrt{-G}} 
\frac{-\partial_r \Omega}{(E - \Omega L)^2} 
\int_{R_{\rm ISCO}}^{r} (E - \Omega L) 
(\partial_\rho L) d\rho \, , \nonumber\\
\ee
$\dot{M}$ is the mass accretion rate. $E$, $L$, and $\Omega$ are, respectively, the specific energy, the axial component of the specific angular momentum, and the orbital frequency of equatorial circular orbits, while $G$ is the determinant of the near equatorial plane metric.

\begin{figure*}[t]
\begin{center}
\includegraphics[type=pdf,ext=.pdf,read=.pdf,width=8.5cm]{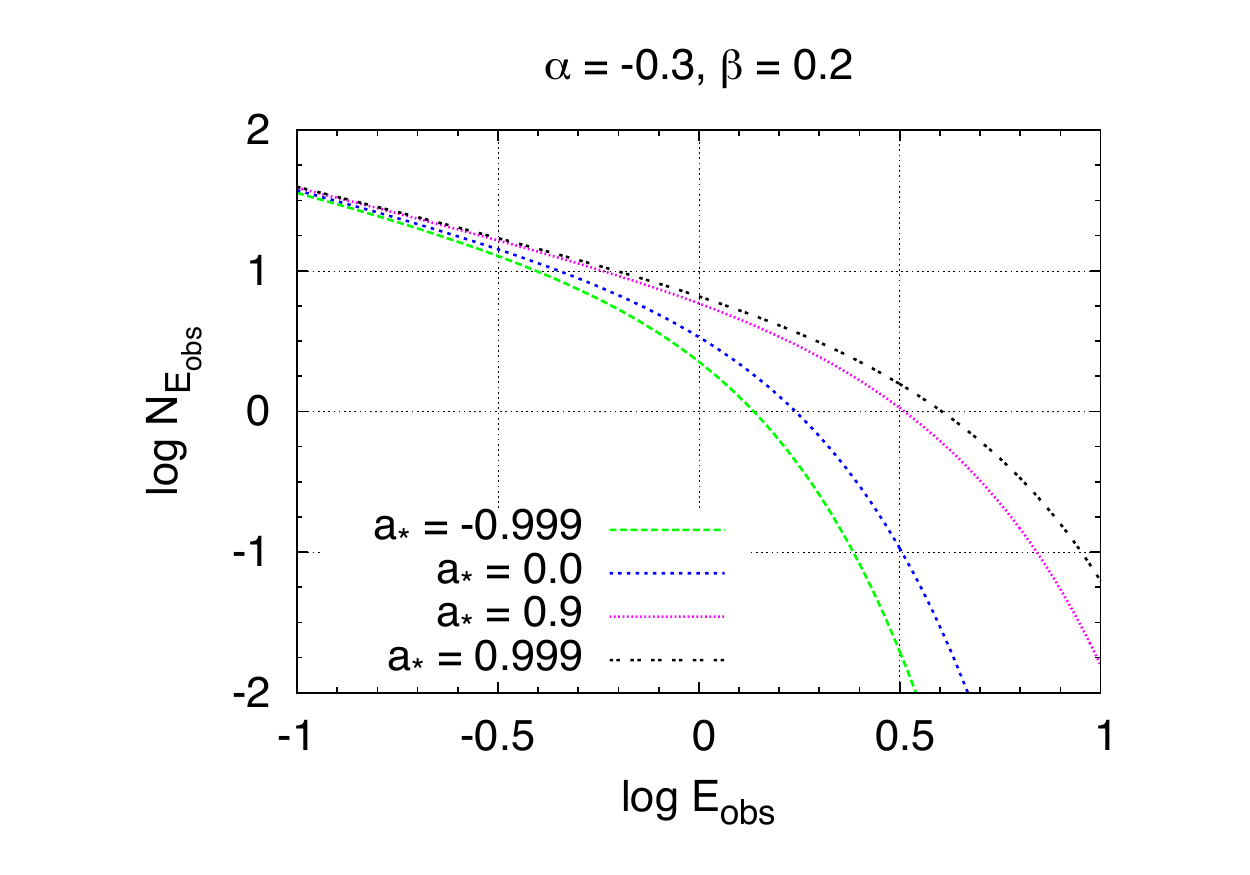}
\hspace{-0.3cm}
\includegraphics[type=pdf,ext=.pdf,read=.pdf,width=8.5cm]{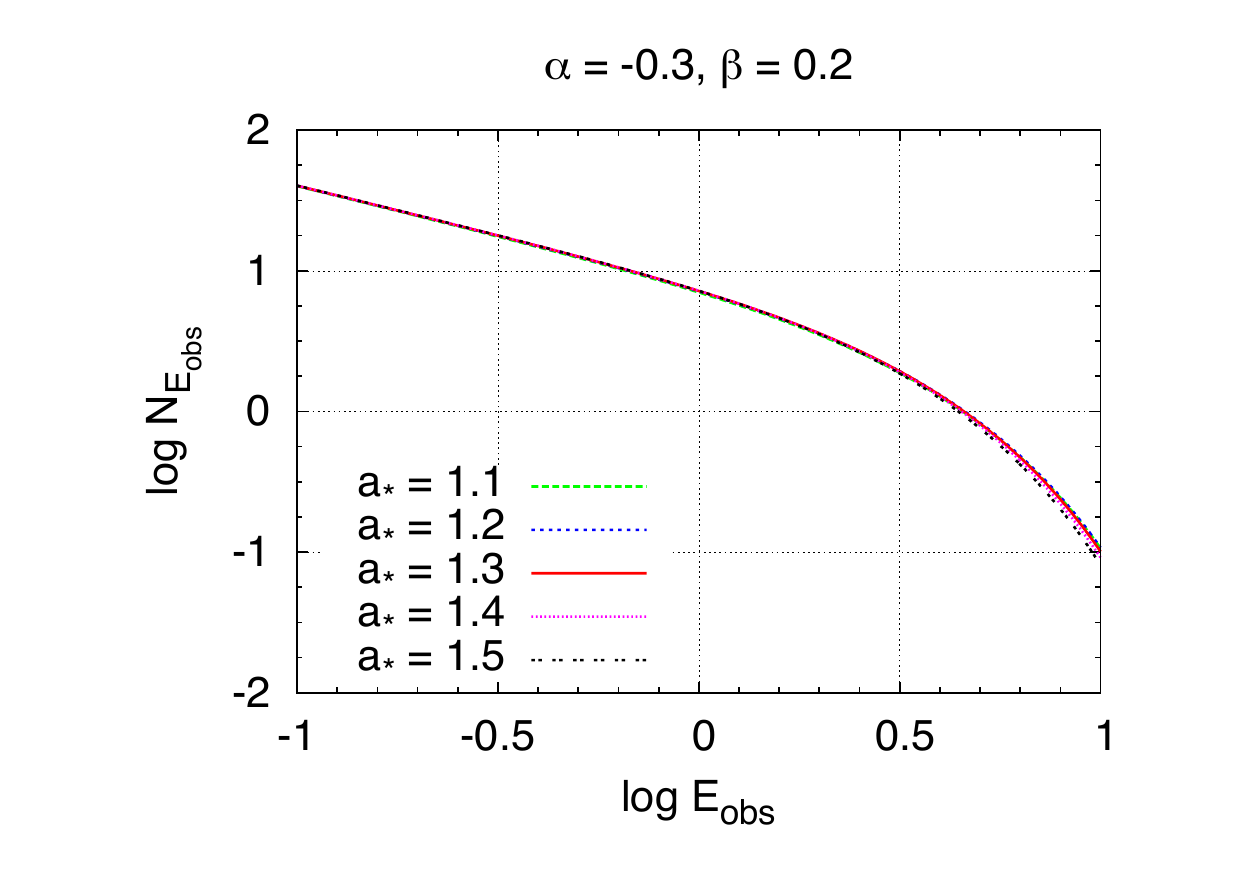}
\end{center}
\vspace{-0.3cm}
\caption{Thermal spectrum of thin disks in non-Kerr spacetimes with $\alpha = -0.3$ and $\beta = 0.2$. The other model parameters are the same as in Fig.~\ref{fig-cfm1}. These spectra can be understood in terms of the Novikov-Thorne radiative efficiency. See the text for more details.}
\label{fig-cfm2}
\end{figure*}

Observations measure the flux at the distance $r_{\rm o}$ of the observer. The impact of our deformation parameters on the thermal spectrum of thin disks is illustrated in Figs.~\ref{fig-cfm1} and \ref{fig-cfm2}. The results could have been imagined from the contour levels of the Novikov-Thorne radiative efficiency $\eta_{\rm NT}$~\cite{tt4}. The shape of the spectrum is simple, as it is just a multi-blackbody spectrum because the disk radiates as a blackbody locally and then one has to integrate radially. The high energy cut-off of the spectrum is determined by the Novikov-Thorne radiative efficiency, and therefore the measurement of the thermal component of the disk roughly corresponds to the measurement of $\eta_{\rm NT}$. We cannot distinguish objects with the same $\eta_{\rm NT}$ from the sole observation of the disk's thermal spectrum, and therefore objects on the same contour level of $\eta_{\rm NT}$ have substantially the same spectrum. We note that in our figures we have considered even objects with spin parameter larger than 1. In the Kerr metric, for $|a_*| > 1$ there is no black hole and there are reasons to ignore these objects (see e.g. the discussion in Ref.~\cite{rev1}). In our case, these objects are black holes and they cannot be excluded {\it a priori}.

\section{Concluding remarks}

In this work, we have discussed a simple parametrization to describe possible deviations from the Kerr metric and test astrophysical black hole candidates. Our metric is suitable for the numerical calculations required to evaluate the electromagnetic spectrum of thin disks, in particular the thermal spectrum and the iron line profile. These are today the two leading techniques to probe the spacetime geometry around black hole candidates~\cite{bjs15}.

The continuum-fitting method is normally used for stellar-mass black hole candidates only, because the temperature of a Novikov-Thorne disk scales as $M^{-0.25}$: for $M \approx 10$~$M_\odot$, the spectrum is in the soft X-ray band, while for supermassive black hole candidates it is in the UV/optical bands, where dust absorption makes an accurate measurement impossible. Stronger constraints on possible deviations from the Kerr metric can be obtained when the inner edge of the disk is very close to the compact object~\cite{ooo}. The best target may be the black hole binary Cygnus~X-1~\cite{o1}.

The iron line method can be used for both stellar-mass and supermassive black hole candidates, because the measurement does not directly depend on the mass of the object. Stellar-mass black hole candidates have the advantage to be brighter, so the photon count number in the iron line is high enough, which is not the case in AGN data. However, the spectrum of black hole binaries is more difficult to model (mainly because of the higher temperature of the disk) and the low energy tail of the iron line overlaps with the thermal component of the disk. At the moment it is not clear if the best targets to test the Kerr metric with the iron line are stellar-mass or supermassive black hole candidates. Even in this case, stronger constraints can be obtained if the inner part of the accretion disk is very close to the central object. The sources should also be sufficiently reflection-dominated, to have a stronger contrast between the primary and the reflection component. Cygnus~X-1 should again be one of the most promising targets in the case of black hole binaries~\cite{o2}. For AGN, current observations suggest that good targets would be NGC1365~\cite{o3} and 1H0707-495~\cite{o4}, as both sources are bright, their reflection component is strong, and the inner edge of the disk seems to be very close to the black hole candidate.

In the analysis of real data, it is usually necessary to compute many spectra to fit the model parameters. In the Kerr case, it is possible to exploit a number of nice properties, with the results that numerical calculations can be fast and accurate. These properties are usually absent in non-Kerr metrics, and the calculation times become too long. To have a rough idea, the ray tracing calculations for one spectrum in the Kerr metric take about 5~minutes with a standard computer, so the calculation of a grid of, say, 40~spins and 20~angles takes something like 3~days. If the calculations are done by solving the geodesic equations, the computation time increases by about an order of magnitude is we want the same calculation accuracy. Moreover, we do not have a 2D grid any more, but a 3D grid if we consider one deformation parameter, or a 4D grid if we have two deformation parameters at the same time. Any parametrization has its own advantages and disadvantages, and in any case we cannot pretend to test black hole candidates with the most general black hole solution, because this would require an infinite number of deformation parameters and even in the presence of high quality data it is impossible to constrain several deformation parameters at the same time. Bearing in mind that any parametrization is necessary {\it ad hoc} and subject to criticisms, in our case we have looked for something suitable to analize X-ray data.


\begin{acknowledgments}
This work was supported by the NSFC grant No.~11305038, the Shanghai Municipal Education Commission grant No.~14ZZ001, the Thousand Young Talents Program, and Fudan University. C.B. acknowledges also support from the Alexander von Humboldt Foundation. M.G.-N. acknowledges also support from China Scholarship Council (CSC), grant No.~2014GXZY08, and thanks the School of Astronomy at the Institute for Research in Fundamental Sciences (IPM), Tehran, where part of this work was done.
\end{acknowledgments}


\appendix

\section{Invariants}

With the choice of $m_1$ and $m_2$ in Eq.~(\ref{eq-Mab}), we compute the invariants $R$, $R^{\mu\nu} R_{\mu\nu}$, and $R^{\mu\nu\rho\sigma} R_{\mu\nu\rho\sigma}$ to verify that the metric is regular outside the event horizon. These three quantities indeed only diverge at $r=0$. The scalar curvature $R$ is
\be
R = - \frac{4 M^3}{r^3} \frac{1}{\Sigma} \Bigg(\alpha + 3 \beta \frac{M}{r}\Bigg) \, .
\ee
The quare of the Ricci tensor is given by
\begin{widetext}
\be
R^{\mu\nu} R_{\mu\nu} &=& \frac{8 M^6}{r^2} \frac{1}{\Sigma^4}
\Bigg(13 \alpha^2 + 48 \alpha \beta \frac{M}{r} 
+ 6 \alpha^2 \frac{a^2 x^2}{r^2} + 45 \beta^2 \frac{M^2}{r^2}
+ 30 \alpha \beta \frac{a^2 M x^2}{r^3}
+ \alpha^2 \frac{a^4 x^4}{r^4} + 36 \beta^2 \frac{a^2 M^2 x^2}{r^4}
\nonumber\\ && \hspace{1.5cm}
+ 6 \alpha \beta \frac{a^4 M x^4}{r^5} + 9 \beta^2 \frac{a^4 M^2 x^4}{r^6}\Bigg) \, ,
\ee
where $x = \cos\theta$. The Kretschmann scalar is
\be
R^{\mu\nu\rho\sigma} R_{\mu\nu\rho\sigma} &=& \frac{16 M^2 r^6}{\Sigma^6}
\Bigg(3 - 45 \frac{a^2 x^2}{r^2} +20 \alpha \frac{M^2}{r^2} 
+ 30 \beta \frac{M^3}{r^3}
+ 45 \frac{a^4 x^4}{r^4} - 152 \alpha \frac{a^4 M^2 x^2}{r^4} + 46 \alpha^2 \frac{M^4}{r^4}
\nonumber\\ && \hspace{1.7cm} 
- 186 \beta \frac{a^2 M^3 x^2}{r^5} + 146 \alpha \beta \frac{M^5}{r^5}
- 3 \frac{a^6 x^6}{r^6} + 20 \alpha \frac{a^4 M^2 x^4}{r^6} - 51 \alpha^2 \frac{a^2 M^4 x^2}{r^6} + 117 \beta^2 \frac{M^6}{r^6}
\nonumber\\ && \hspace{1.7cm}
- 30 \beta \frac{a^4 M^3 x^4}{r^7} - 32 \alpha \beta \frac{a^2 M^5 x^2}{r^7}
+ \alpha^2 \frac{a^4 M^4 x^4}{r^8} + 66 \beta^2 \frac{a^2 M^6 x^2}{r^8}
+ 32 \alpha \beta \frac{a^4 M^5 x^4}{r^9} - 6 \beta \frac{a^6 M^3 x^6}{r^9}
\nonumber\\ && \hspace{1.7cm}
+ 3 \alpha^2 \frac{a^6 M^4 x^6}{r^{10}} + 78 \beta^2 \frac{a^4 M^6 x^4}{r^{10}}
+ 24 \alpha \beta \frac{a^6 M^5 x^6}{r^{11}} 
+ \alpha^2 \frac{a^8 M^4 x^8}{r^{12}} + 42 \beta^2 \frac{a^6 M^6 x^6}{r^{12}}
\nonumber\\ && \hspace{1.7cm}
+ 6 \alpha \beta \frac{a^8 M^5 x^8}{r^{13}}
+ 9 \beta^2 \frac{a^8 M^6 x^8}{r^{14}} \Bigg) \, .
\ee
\end{widetext}



\begin{thebibliography}{99}

\bibitem{rr} 
  C.~E.~Rhoades, Jr. and R.~Ruffini,
  Phys.\ Rev.\ Lett.\  {\bf 32}, 324 (1974).

\bibitem{maoz} 
  E.~Maoz,
  Astrophys.\ J.\  {\bf 494}, L181 (1998)
  [astro-ph/9710309].

\bibitem{h1} 
  J.~E.~McClintock, R.~Narayan and G.~B.~Rybicki,
  Astrophys.\ J.\  {\bf 615}, 402 (2004)
  [astro-ph/0403251].
  
\bibitem{h2} 
  A.~E.~Broderick, A.~Loeb and R.~Narayan,
  Astrophys.\ J.\  {\bf 701}, 1357 (2009)
  [arXiv:0903.1105 [astro-ph.HE]].  

\bibitem{k1} 
  R.~H.~Price,
  Phys.\ Rev.\ D {\bf 5}, 2419 (1972).

\bibitem{k2} 
  C.~Bambi, A.~D.~Dolgov and A.~A.~Petrov,
  JCAP {\bf 0909}, 013 (2009)
  [arXiv:0806.3440 [astro-ph]].

\bibitem{k3} 
  C.~Bambi, D.~Malafarina and N.~Tsukamoto,
  Phys.\ Rev.\ D {\bf 89}, 127302 (2014)
  [arXiv:1406.2181 [gr-qc]].

\bibitem{rev1} 
  C.~Bambi,
  arXiv:1509.03884 [gr-qc].

\bibitem{rev2} 
  C.~Bambi,
  Mod.\ Phys.\ Lett.\ A {\bf 26}, 2453 (2011)
  [arXiv:1109.4256 [gr-qc]].

\bibitem{ppn68} 
  K.~Nordtvedt,
  Phys.\ Rev.\  {\bf 169}, 1017 (1968).

\bibitem{p1} 
  T.~Johannsen and D.~Psaltis,
  Phys.\ Rev.\ D {\bf 83}, 124015 (2011)
  [arXiv:1105.3191 [gr-qc]].

\bibitem{p2} 
  V.~Cardoso, P.~Pani and J.~Rico,
  Phys.\ Rev.\ D {\bf 89}, 064007 (2014)
  [arXiv:1401.0528 [gr-qc]]. 
  
\bibitem{p3} 
  K.~Glampedakis and S.~Babak,
  Class.\ Quant.\ Grav.\  {\bf 23}, 4167 (2006)
  [gr-qc/0510057].  
  
\bibitem{p4} 
  S.~Vigeland, N.~Yunes and L.~Stein,
  Phys.\ Rev.\ D {\bf 83}, 104027 (2011)
  [arXiv:1102.3706 [gr-qc]].  
  
\bibitem{p5} 
  T.~Johannsen,
  Phys.\ Rev.\ D {\bf 88}, 044002 (2013)
  [arXiv:1501.02809 [gr-qc]].  
  
\bibitem{p6} 
  L.~Rezzolla and A.~Zhidenko,
  Phys.\ Rev.\ D {\bf 90}, 084009 (2014)
  [arXiv:1407.3086 [gr-qc]].  
  
\bibitem{tt1} 
  C.~Bambi and E.~Barausse,
  Astrophys.\ J.\  {\bf 731}, 121 (2011)
  [arXiv:1012.2007 [gr-qc]].  
  
\bibitem{tt2} 
  C.~Bambi,
  Astrophys.\ J.\  {\bf 761}, 174 (2012)
  [arXiv:1210.5679 [gr-qc]].  
  
\bibitem{tt3} 
  C.~Bambi,
  Phys.\ Rev.\ D {\bf 87}, 023007 (2013)
  [arXiv:1211.2513 [gr-qc]].  
  
\bibitem{tt4} 
  L.~Kong, Z.~Li and C.~Bambi,
  Astrophys.\ J.\  {\bf 797}, 78 (2014)
  [arXiv:1405.1508 [gr-qc]].  
  
\bibitem{tt5}  
  J.~Jiang, C.~Bambi and J.~F.~Steiner,
  Astrophys.\ J.\  {\bf 811}, 130 (2015)
  [arXiv:1504.01970 [gr-qc]].
  
\bibitem{tt6} 
  T.~Johannsen and D.~Psaltis,
  Astrophys.\ J.\  {\bf 773}, 57 (2013)
  [arXiv:1202.6069 [astro-ph.HE]].  
  
\bibitem{tt7} 
  T.~Johannsen,
  Phys.\ Rev.\ D {\bf 90}, 064002 (2014)
  [arXiv:1501.02815 [astro-ph.HE]].  
  
\bibitem{bjs15} 
  C.~Bambi, J.~Jiang and J.~F.~Steiner,
  arXiv:1511.07587 [gr-qc].  
  
\bibitem{nc1} 
  P.~Nicolini, A.~Smailagic and E.~Spallucci,
  Phys.\ Lett.\ B {\bf 632}, 547 (2006)
  [gr-qc/0510112].  

\bibitem{nc2} 
  S.~Ansoldi, P.~Nicolini, A.~Smailagic and E.~Spallucci,
  Phys.\ Lett.\ B {\bf 645}, 261 (2007)
  [gr-qc/0612035].
  
\bibitem{nc3} 
  A.~Smailagic and E.~Spallucci,
  Phys.\ Lett.\ B {\bf 688}, 82 (2010)
  [arXiv:1003.3918 [hep-th]].  
  
\bibitem{nc4} 
  L.~Modesto and P.~Nicolini,
  Phys.\ Rev.\ D {\bf 82}, 104035 (2010)
  [arXiv:1005.5605 [gr-qc]].  

\bibitem{wngt}
  Y.~Zhang, Y.~Zhu, L.~Modesto and C.~Bambi,
  Eur.\ Phys.\ J.\ C {\bf 75}, 96 (2015)
  [arXiv:1404.4770 [gr-qc]].
  
\bibitem{bardeen} 
  J.~M.~Bardeen,
  in {\it Conference Proceedings of GR5} (Tbilisi, USSR, 1968), p. 174.  
  
\bibitem{regular} 
  C.~Bambi and L.~Modesto,
  Phys.\ Lett.\ B {\bf 721}, 329 (2013)
  [arXiv:1302.6075 [gr-qc]].      
  
\bibitem{bardeen2} 
  E.~Ayon-Beato and A.~Garcia,
  Phys.\ Lett.\ B {\bf 493}, 149 (2000)
  [gr-qc/0009077].    
  
\bibitem{will} 
  C.~M.~Will,
  Living Rev.\ Rel.\  {\bf 9}, 3 (2006)
  [gr-qc/0510072].  
  
\bibitem{nt-model}
  I.~D.~Novikov, K.~S.~Thorne,
  ``Astrophysics of Black Holes'' in {\it Black Holes}, 
  edited by C.~De~Witt and B.~De~Witt
  (Gordon and Breach, New York, US, 1973), pp. 343-450.  
  
\bibitem{c75} 
  C.~T.~Cunningham,
  Astrophys.\ J.\  {\bf 202}, 788 (1975).  
  
\bibitem{srr} 
  R.~Speith, H.~Riffert and H.~Ruder,
  Comput.\ Phys.\ Commun.\  {\bf 88}, 109 (1995).   
  
\bibitem{teu72} 
  J.~M.~Bardeen, W.~H.~Press and S.~A.~Teukolsky,
  Astrophys.\ J.\  {\bf 178}, 347 (1972).  
  
\bibitem{bb2} 
  C.~Bambi and E.~Barausse,
  Phys.\ Rev.\ D {\bf 84}, 084034 (2011)
  [arXiv:1108.4740 [gr-qc]].  
  
\bibitem{chandra} 
  S.~Chandrasekhar,
  {\it The Mathematical Theory of Black Holes}
  (Clarendon Press, Oxford, UK, 1998).    
  
\bibitem{lixin} 
  L.~X.~Li, E.~R.~Zimmerman, R.~Narayan and J.~E.~McClintock,
  Astrophys.\ J.\ Suppl.\  {\bf 157}, 335 (2005)
  [astro-ph/0411583].  
  
\bibitem{naoc} 
  F.~Zhang, Y.~Lu and Q.~Yu,
  Astrophys.\ J.\  {\bf 809}, 127 (2015)
  [arXiv:1508.06293 [astro-ph.HE]].   
  
\bibitem{2011} 
  V.~Z.~Enolski, E.~Hackmann, V.~Kagramanova, J.~Kunz and C.~Lammerzahl,
  J.\ Geom.\ Phys.\  {\bf 61}, 899 (2011)
  [arXiv:1011.6459 [gr-qc]].   
  
\bibitem{ooo} 
  C.~Bambi,
  Phys.\ Lett.\ B {\bf 730}, 59 (2014)
  [arXiv:1401.4640 [gr-qc]].  
  
\bibitem{o1}
  L.~Gou {\it et al.},
  Astrophys.\ J.\  {\bf 790}, no. 1, 29 (2014)
  [arXiv:1308.4760 [astro-ph.HE]].

\bibitem{o2}  
  A.~C.~Fabian {\it et al.},
  Mon.\ Not.\ Roy.\ Astron.\ Soc.\  {\bf 424}, 217 (2012)
  [arXiv:1204.5854 [astro-ph.HE]].     
  
\bibitem{o3} 
  D.~J.~Walton {\it et al.},
  Astrophys.\ J.\  {\bf 788}, 76 (2014)
  [arXiv:1404.5620 [astro-ph.HE]].  
  
\bibitem{o4} 
  A.~Zoghbi, A.~Fabian, P.~Uttley, G.~Miniutti, L.~Gallo, C.~Reynolds, J.~Miller and G.~Ponti,
  Mon.\ Not.\ Roy.\ Astron.\ Soc.\  {\bf 401}, 2419 (2010)
  [arXiv:0910.0367 [astro-ph.HE]].  

\end{thebibliography}
\end{document}